\documentclass[revtex4-1,twocolumn,useAMS,usenatbib]{mn2e}
\usepackage{times}
\usepackage{graphicx,natbib}
\usepackage{ulem} 
\usepackage{url}
\usepackage{color}
\usepackage{rotating}
\def\kmsMpc{{\rm km/s/Mpc}}

\newcommand{\degree}{\hbox{$^\circ$}}

\newcommand{\fsky}{f_{\rm sky}}

\newcommand{\hinv}{h^{-1}}

\newcommand{\zphot}{z_{\rm phot}}
\newcommand{\zspec}{z_{\rm spec}}
\newcommand{\ztrue}{z_{\rm true}}
\newcommand{\enne}{e_{\rm NNE}}
\newcommand{\qest}{Q_{\rm est}}
\newcommand{\ssrt}{{\rm SSR}_{\rm T}}
\newcommand{\ssro}{{\rm SSR}_{\rm O}}

\newcommand{\Msun}{M_{\odot}}

\newcommand{\nwei}{N(z)_{\rm wei}}

\newcommand{\pzs}{P(z_s|z_p)}
\newcommand{\pztrue}{P(z_{\rm true}|z_p)}
\newcommand{\pzst}{P(z_{\rm spec}|z_{\rm true})}

\newcommand{\qcrit}{$\rm Q_{\rm crit}$}

\newcommand{\sse}{\sigma_{68}}

\begin{document}    

\title{Spectroscopic failures in photometric redshift calibration: cosmological biases and survey requirements}

\author[Cunha et al.]{Carlos E. Cunha$^{1,2}$\thanks{{\texttt ccunha@stanford.edu}},
Dragan Huterer$^{1}$,
Huan Lin$^{3}$,
Michael T. Busha$^{2,4}$,
Risa H. Wechsler$^{2,5}$\vspace{0.3cm}
\\
${}^{1}$Department of Physics, University of Michigan, 450 Church St, Ann Arbor, MI 48109-1040\\
${}^{2}$Kavli Institute for Particle Astrophysics and Cosmology 452 Lomita Mall, Stanford University, Stanford, CA, 94305\\
${}^{3}$Center for Particle Astrophysics, Fermi National Accelerator Laboratory, Batavia, IL 60510 \\
${}^{4}$Institute for Theoretical Physics, University of Zurich, 8057 Zurich, Switzerland\\
${}^{5}$Department of Physics, Stanford University, Stanford, CA, 94305,\\
SLAC National Accelerator Laboratory, 2575 Sand Hill Rd., MS 29, Menlo Park, CA, 94025
}

\date{\today}

\maketitle

\begin{abstract}

We use N-body-spectro-photometric simulations to investigate the impact of
incompleteness and incorrect redshifts in spectroscopic surveys to photometric
redshift training and calibration and the resulting effects on cosmological
parameter estimation from weak lensing shear-shear correlations.  The
photometry of the simulations is modeled after the upcoming Dark Energy Survey
and the spectroscopy is based on a low/intermediate resolution spectrograph
with wavelength coverage of $5500 {\rm \AA} <\lambda< 9500 {\rm \AA}$.  The principal
systematic errors that such a spectroscopic follow-up encounters are
incompleteness (inability to obtain spectroscopic redshifts for certain
galaxies) and wrong redshifts.  Encouragingly, we find that a neural
network-based approach can effectively describe the spectroscopic
incompleteness in terms of the galaxies' colors, so that the spectroscopic
selection can be applied to the photometric sample.  Hence, we find that
spectroscopic incompleteness yields no appreciable biases to cosmology,
although the statistical constraints degrade somewhat because the photometric
survey has to be culled to match the spectroscopic selection.  Unfortunately,
wrong redshifts have a more severe impact: the cosmological biases are
intolerable if more than a percent of the spectroscopic redshifts are incorrect.
Moreover, we find that incorrect redshifts can substantially degrade the
perceived accuracy of training set based photo-z estimators,
though the actual accuracy is virtually unaffected.  The main problem is the
difficulty of obtaining redshifts, either spectroscopically or
photometrically, for objects at $z>1.3$.  We discuss several approaches for
reducing the cosmological biases, in particular finding that photo-z error
estimators can reduce biases appreciably when the photo-z 
errors are correlated with the spectroscopic failures, but not otherwise.

\end{abstract}

\section{Introduction}\label{sec:intro}

Large-scale structure surveys benefit enormously from the information about
galaxy redshifts. The redshift information reveals the third spatial dimension 
of a galaxy survey, enabling a much more accurate mapping of the expansion and 
growth history of the Universe relative to the case when only angular information 
is available. 
Unfortunately, obtaining spectroscopic redshifts for all galaxies is typically 
impossible in wide-field imaging surveys due to the large number 
($\sim 10^8$-$10^9$) of  galaxies and the high cost of spectroscopy, 
especially for the high-redshift galaxies. 
To circumvent this problem, the current approach in the community is to estimate
redshifts using photometric measurements, i.e. fluxes from a few broad band filters.
These redshift estimates are known as photometric redshifts, or photo-zs, and 
are necessarily coarser than spectroscopic redshifts. 
Because of the intrinsically large errors, photo-zs typically cannot be used 
directly for cosmological analysis, unless the photo-z error distributions 
can be quantified precisely.

The standard approach to quantify, or calibrate, the photo-z error distributions
is to use a small subsample of galaxies with known redshifts.
As discussed in detail in \cite{cun12}, spectroscopic samples used to
train photo-zs (cf. Sec. \ref{sec:train}) need to be locally (in the 
space of observables) representative subsamples of the photometric samples.
For calibration of the photo-z {\it error distributions}, however, the 
spectroscopic sample must be globally representative.  
More specifically, the ideal spectroscopic survey 
should satisfy the following properties:

\begin{itemize}

\item {\it Large area:}
  A spectroscopic survey needs to span a large area to beat down sample
  variance, and has to have tens of thousands of galaxies to beat down
  shot-noise in the photo-z error calibration \citep{cun12}.  In addition, the
  spectroscopic sample needs to be imaged under conditions
  that faithfully reproduce the variations in
  the full photometric sample \citep[see e.g.][]{nak12}.
  Note that requirements
might   be alleviated with a correction to the individual galaxy redshift 
  likelihoods \citep{bor10,bor12}.  
  In the context of dark energy parameter constraints, however, a full 
  analysis that goes beyond the overall redshift distribution and involves 
  the full error matrix $P(z_s|z_p)$ is required \citep{BH10, Hearin}.

\item {\it High completeness:} The spectroscopic survey needs to span the same
  range of redshifts, galaxy types, and other observational selection 
  parameters as the photometric survey.
  When this is not possible, we say that the survey is {\it incomplete}. 
  In that case, the photometric survey has to be culled to
  ensure both surveys have matching selections.  
  Alternatively, the galaxies in the spectroscopic survey can be weighted
  so as to reproduce the statistical properties of the photometric sample.
  Achieving high completeness in faint spectroscopic surveys is a major 
  challenge.

\item {\it Few wrong redshifts:} We show in this paper that spectroscopic surveys
  need to have extremely accurate redshifts. As shown by many authors 
  \citep[e.g. ][]{ma06,hut06,Amara_Refregier_optimal,Abdalla08,Bernstein_Ma,Kitching_sys,Hearin}
  the photo-z calibration requires exquisite knowledge of the photo-z error distribution.  
  Errors in the  spectroscopic redshifts impair the characterization of the photo-z 
  errors and severely degrade our ability to extract cosmological constraints from
  photometric surveys. 
\end{itemize}

For fixed observing resources, there is a
conflict between accurate redshifts and completeness goals: as we stretch the
observational limits (i.e.\ by observing very faint galaxies) to sample
redshifts that would mimic the distribution of the photometric sample, we
increase the fraction of incorrect spectroscopic redshifts.
As we will show, redshift accuracy is more important for the upcoming surveys.
Similarly, for existing instruments, the requirements on large
  area and number of objects implies that redshifts will have to be
  obtained with low- and mid-resolution spectroscopy. 
With low resolution, is it harder to get redshifts for galaxies with
narrow emission lines that fall in between the night sky lines. 
As a result, it is harder to reach high completeness with realistic 
integration times and the selection functions of low-resolution 
surveys are harder to quantify. 
Until the next generation of spectroscopic instruments becomes 
available, existing and upcoming photometric surveys will have to
learn to deal with very incomplete spectroscopic samples for photo-z calibration.

The purpose of this paper is to assess the impact of spectroscopic selection,
i.e. completeness and accuracy, on the training and calibration of photometric
redshifts and the resulting impact on cosmological constraints derived from
weak lensing shear-shear correlations.  To achieve this goal, we combine
N-body, photometric and spectroscopic simulations patterned after the proposed
characteristics of the Dark Energy Survey (DES) and expected spectroscopic
follow-up.  We then propagate the errors due to imperfect photo-z
calibration on the cosmological parameter constraints inferred from the weak
gravitational lensing power spectrum observations forecasted for the DES.

The paper is organized as follows.  In Sec. \ref{sec:specintro} we
 provide a pedagogical introduction to the main issues driving 
completeness and accuracy of a spectroscopic sample.
In Sec. \ref{sec:data} we briefly describe the simulated catalogs we use, 
leaving the details of the catalog generation to Appendix \ref{app:sims}.  
In Sec. \ref{sec:meth} we give a step-by-step guide describing how 
we go from the simulated data to the cosmological constraints, detailing 
the methods used at each step.
Results are presented in Secs. \ref{sec:results1} and \ref{sec:results2}.
We discuss the robustness of our assumptions in Sec. \ref{sec:robust} and
the implications of our findings for spectroscopic survey design in Sec. \ref{sec:design}. 
We present conclusions in Sec. \ref{sec:concl}.

\begin{figure*}
\centering
   \includegraphics[scale=0.4]{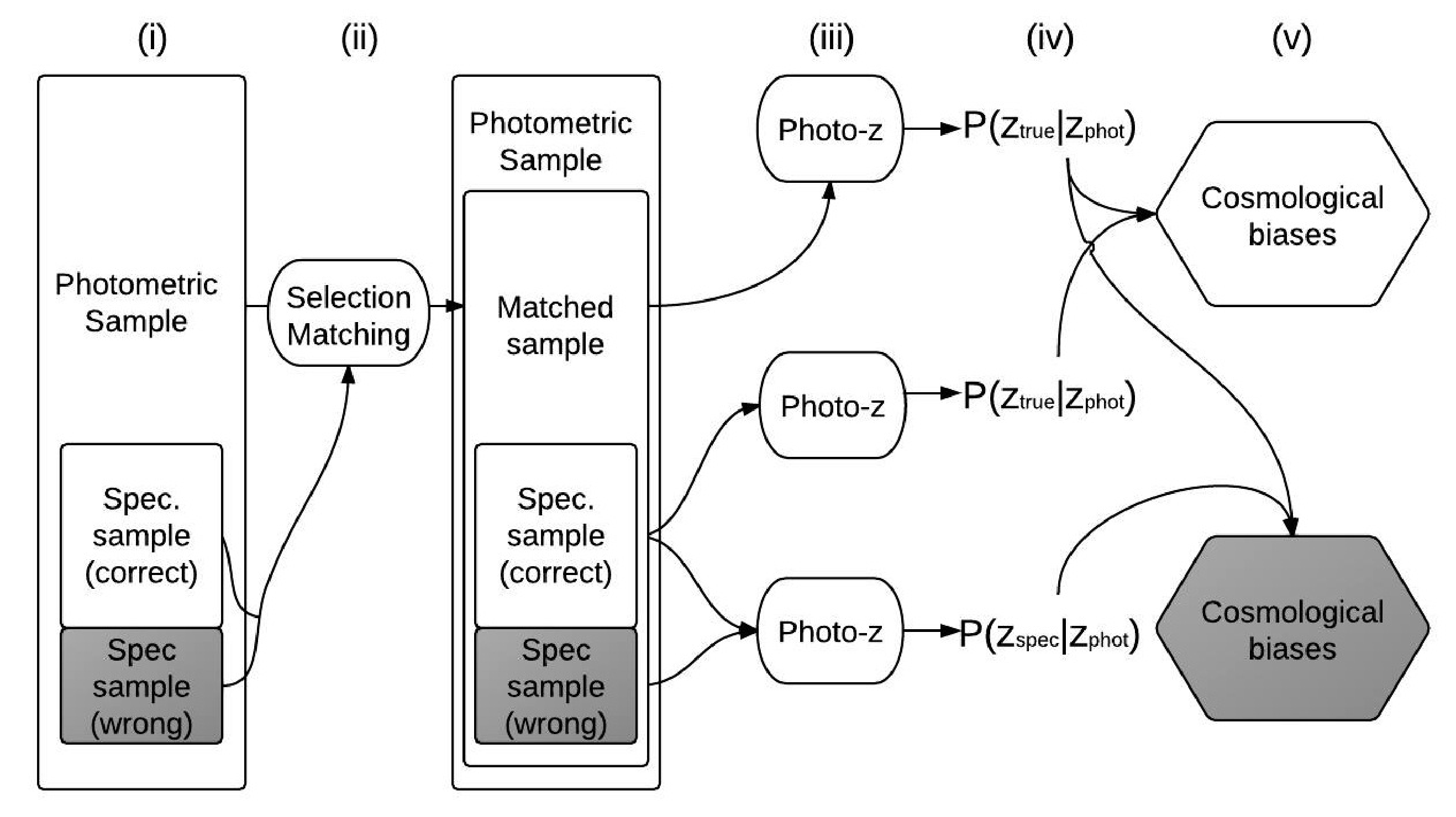}
\caption{Flowchart describing our step-by-step procedure to go from the simulated 
observations to cosmological biases.}
\label{fig:chart}
\end{figure*}

\section{Basics of low-resolution spectroscopy}\label{sec:specintro}

In this section we provide a brief pedagogical overview of issues in 
spectroscopy, targeted to theorists.

\subsection{Key parameters of spectroscopic surveys}\label{sec:keyspec}

Spectroscopic redshifts are often derived by 
cross-correlating a library of galaxy templates with observed (or simulated) spectra.
For fixed observing conditions (and in the absence of instrumental systematic effects), 
three main items determine the quality of the estimated spectroscopic redshifts:

\begin{enumerate}
\item {\it Spectral coverage:} The wavelength range covered by the
  spectrograph needs to bracket a few significant spectral features.  As shown
  in the bottom plot of Fig. \ref{fig:pipe}, for our simulation the coverage
  is roughly from $5500{\rm \AA}$ to $9500{\rm \AA}$, with decreasing sensitivity at
  longer wavelengths.
\item {\it Integration time:} The faintest galaxies detectable by upcoming optical
  surveys can be a few orders of magnitude fainter than the atmospheric emission.
  Thus, significant integration times, as well as careful subtraction of the sky 
  background, are needed to obtain secure redshift measurements.
\item {\it Cross-correlation templates:} Having an accurate and representative
  set of galaxy spectral distribution templates is important in deriving
  accurate redshifts and associated uncertainties.  As we discuss in the next
  section, this is particularly important for early-type galaxies and galaxies
  at $z > 1.5$ (also known as the {\it redshift desert}) because of the lack
  of strong emission features in the spectrograph window.
\end{enumerate}

\subsection{Principal spectral features}\label{sec:lines}

The two main emission lines used in optical spectroscopy are the [OII]
(singly-ionized oxygen) line at $3727 {\rm \AA}$ and the $H\alpha$ (first transition
in the Balmer series) line at $6563 {\rm \AA}$.  The main absorption feature is the
$4000 {\rm \AA}$ break, caused by a confluence of absorption lines, particularly the
H and K Calcium lines.  In high-resolution spectroscopy, [OII] is the most
important line because it is actually a doublet -- a pair of closely spaced
lines. High-resolution observations - e.g.\ with DEEP2 \citep{new12}, or SDSS
\citep{yor00} - can distinguish the doublet and hence confidently identify [OII].  
Low-resolution observations - e.g.\ as in the VVDS
\citep{lef05} and zCOSMOS surveys \citep{lil07},  
rely on more than one feature.  
The limited spectral range of the instrument sets the regions of redshift 
space where one can confidently identify spectral features.  
In the case of VVDS, for example, there are roughly 5 different redshift regions:
\begin{itemize}
\item $z<0.4$: The $H\alpha$ can be detected, but [OII] cannot. 
  There is risk of confusing $H\alpha$ of a $z<0.4$ galaxy for 
  [OII] emission of a galaxy at $z>0.8$. 
  Fortunately, these galaxies are mostly brighter and thus the $H\alpha$ line 
  combined with less prominent spectral features is often sufficient 
  to estimate a redshift. Similarly, for early type galaxies, the $4000 {\rm \AA}$ break cannot be detected, and one relies on smaller absorption features.

\item $0.4<z<0.6$: Neither [OII] nor $H\alpha$ can be detected. Redshifts have
  to be estimated based on [OIII] and $H\beta$ lines.
\item $0.6<z<0.9$: [OII] and other important lines ([OIII] - $5007 {\rm \AA}$, 
  $H\beta$ - $4861 {\rm \AA}$) are detectable, but get progressively fainter 
  towards higher redshift (due to increasing atmospheric noise and 
  instrumental sensitivity).
\item $0.9<z<1.5$: [OIII] and $H\beta$ are out of the instrument range, 
  but [OII] and the $4000 {\rm \AA}$ break are still detectable.
\item $z>1.5$ (the redshift desert): Only minor features in the spectra
  are available. Visual inspection to reduce incompleteness is
  essential in this range. Potential for wrong redshifts is increased because
  atmospheric emission lines can be mistakenly identified by the algorithm as
  real lines.
\end{itemize}

\subsection{Additional systematics affecting the incompleteness}\label{sec:addsys}

There are a few additional items contributing to the incompleteness of
spectroscopic surveys that are not modeled in our simulations but that exist in real surveys:

\begin{itemize}

\item {\it Fiber collisions and slit overlaps}:  If the angular separation
    between galaxies is too small, one may not simultaneously obtain
    their spectra (without using a multiple pass strategy).  Since
    clustering of galaxies is type dependent, one has to be careful
    that fiber collisions and slit overlaps do not introduce selection
    biases.

\item {\it Optical distortions}:  Geometric distortions due to the
    spectrograph optics may make extraction of spectra and subsequent
    measurement of redshifts more difficult near the edge of the
    instrument field of view.

\item {\it CCD fringing}: Spatial and wavelength 
dependent variations in the pixel response in the red end of the spectrograph.
Fringing hinders measurement of the spectra and redshifts of faint galaxies.


\item {\it Stars and bright galaxies:} Light from nearby stars or bright galaxies 
can contaminate the spectra.

\item {\it Cosmic rays}: Also can contaminate the spectra.

\end{itemize}

Issues such as stars, cosmic rays and edge effects will reduce 
the completeness, more or less randomly, resulting mostly in an increase in the 
shot noise, without galaxy type or redshift dependence.
Note also, that photometric surveys are affected by countless other
selection systematics.
We are only concerned with systematic {\it differences}
between the spectroscopic sample relative to the photometric sample,
however it is defined.

\section{Simulated Data}\label{sec:data}

We use cosmological simulations populated with galaxies and their photometric
properties as described in Appendix \ref{sec:nbody}.
The photometric observations are patterned after the expected sensitivity of
the Dark Energy Survey (DES) and Vista Hemisphere Surveys (VHS), with galaxies 
imaged in the {grizYJHKs} filters over 5100 sq. degrees.
For simplicity, we only use the observations on {\it griz} bands because 
they are imaged for longer periods of time, and hence are useful for all our
sample.
The imaging in these bands is expected to reach $10\sigma$ magnitude limits of
25.2, 24.7, 24.0, and 23.5 in {\it g,r,i} and {\it z}.

For computational efficiency, we select a subsample of approximately 1.3 million 
galaxies, hereafter our { \it photometric sample}, from the total 1 billion galaxies 
present in the simulation.
We apply the same quality cuts as in \cite{cun12}, i.e.\ keep galaxies with $i<24$ 
and at least $5\sigma$ detection in {\it grz}.  
This selection reduces our photometric sample to 726824 galaxies.  

Of this photometric sample, we randomly target a subset of 181892 galaxies, hereafter 
the {\it spectroscopic sample} or {\it training set}, for the spectroscopic analysis.  
The generation of simulated spectra for this subsample is described in the 
Appendix \ref{sec:pipe}.

\section{From the redshifts to cosmology}\label{sec:meth}

In this section, we describe the step-by-step procedure we used for 
converting the simulated observations into cosmological constraints.
The flowchart in Fig. \ref{fig:chart} gives a pictorial version of the
explanation below.

\begin{enumerate}

\item The first step is to estimate spectroscopic redshifts for 
the sample for which we have spectra.  
We use the {\tt rvsao.xcsao} spectral analyzer algorithm described in 
Sec.\ \ref{sec:iraf}.
Not all spectra yield redshifts, and only the redshifts above certain confidence are kept. 
Even so, a fraction of the spectroscopic redshifts is incorrect.

\item The spectroscopic sample can only be used for calibration of the photo-z
error distributions if it is a representative subsample of the photometric
sample. 
Hence, we statistically match spectroscopic and photometric selection 
in one of two ways: by applying the spectroscopic selection to the photometric 
sample with neural networks (cf. Sec. \ref{sec:sel}), or by weighting the 
photometric sample so that its  statistical properties match those of 
the spectroscopic sample (cf. Sec. \ref{sec:weights}).

\item Next, we calculate photo-zs for both the spectroscopic and photometric 
samples, cf. Sec.\ \ref{sec:photoz}.

\item After the matching, we can calculate the photo-z error matrices required
for cosmological analysis.

\item Finally, we estimate fiducial constraints and biases in the cosmological 
parameters forecasted for the DES-type weak gravitational lensing survey.
We break up the tests in two parts. 
In the first case, shown as the transparent hexagon in the flowchart, we only test 
the impact of the selection matching, by using only the correct value for redshifts.
In the second case (gray hexagon), we use the actual value of the spectroscopic 
redshifts - thereby including the small fraction of wrong redshifts.

\end{enumerate}

\begin{figure*}
\centering
   \includegraphics[scale=0.31,angle=-90]{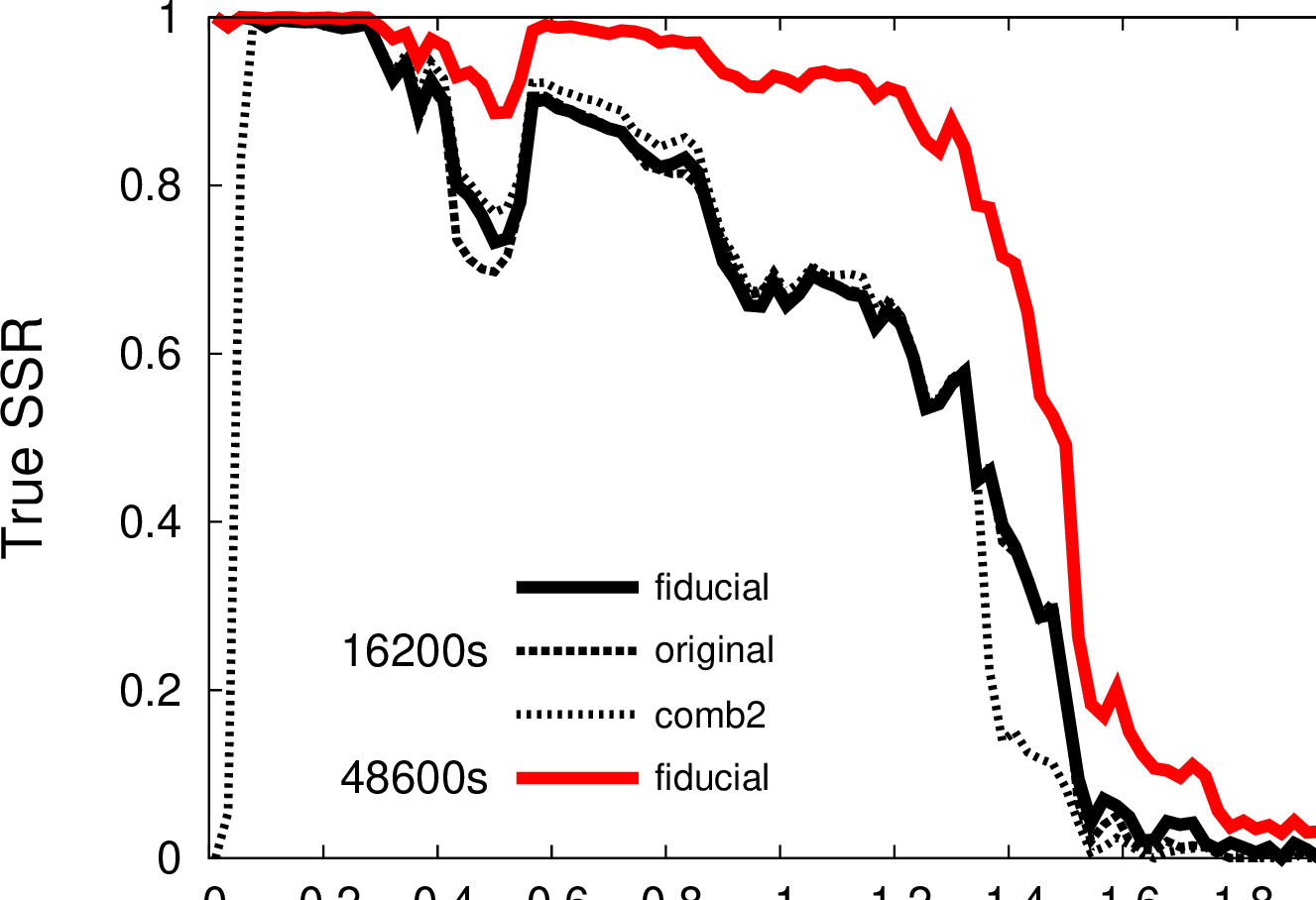}\hspace{0.3cm}
 \includegraphics[scale=0.31,angle=-90]{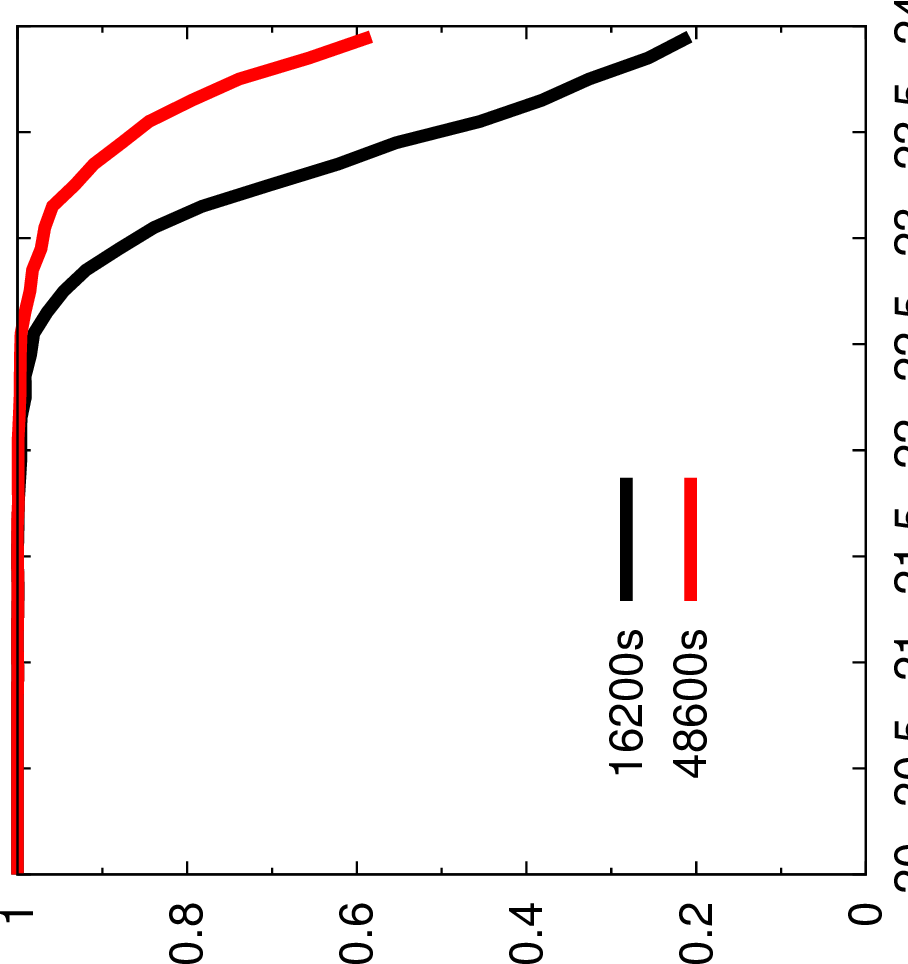}\vspace{0.cm}
  \includegraphics[scale=0.31,angle=-90]{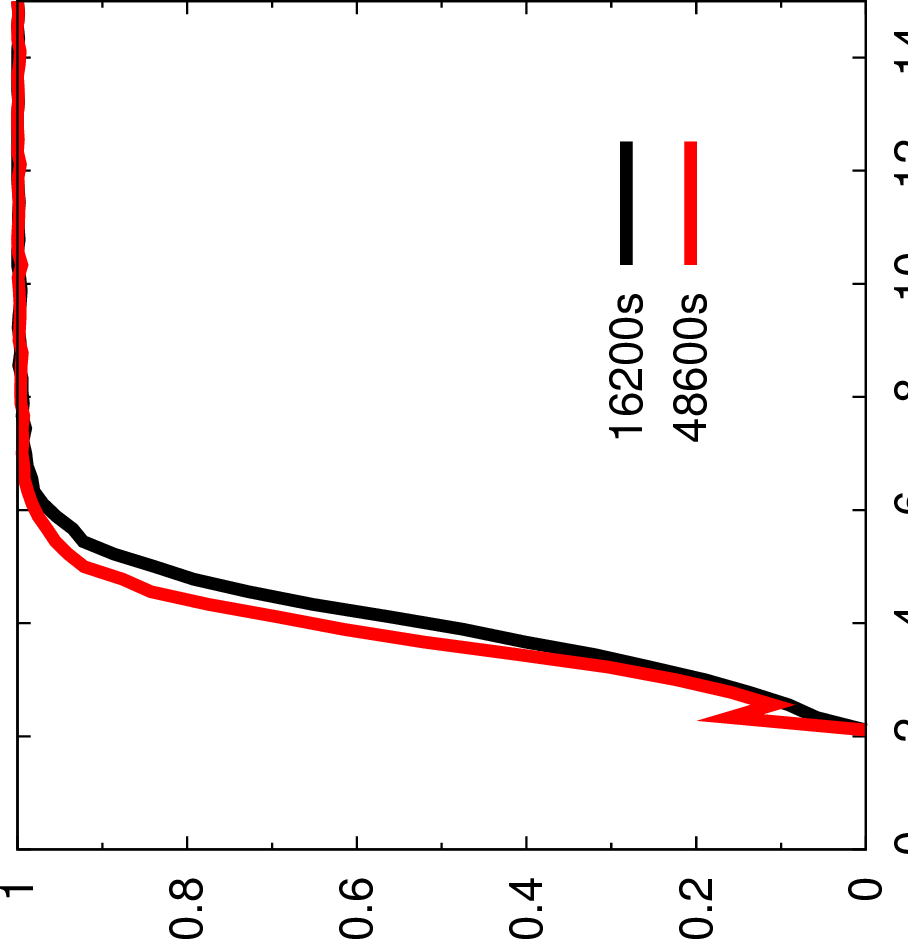}\vspace{0.5cm}

\caption{Left panel: True spectroscopic success rate ($\ssrt$), defined as
  fraction of correct redshifts, as a function of true redshift.  Central
  panel: $\ssrt$ as a function of observed i-band magnitude.  Right panel:
  $\ssrt$ as a function the cross-correlation strength statistic R, which is a
  measure of the redshift confidence.  The black lines assume 16200 secs of
  integration time and the red (gray) lines assume 48600 secs.  The solid, dashed and
  dotted lines correspond to different settings of the spectroscopic pipeline,
  described in Sec. \ref{sec:iraf}.}
\label{fig:ssr}
\end{figure*}

\subsection{Analyzing 1-D spectra}\label{sec:iraf}

Simulating spectroscopic redshift estimation is challenging because real
spectroscopic surveys rely heavily on visual inspection.  
For our forecasts, visual inspection of thousands of spectra would be out
of the question.  Instead, we adopt a more reasonable strategy and apply an
automated pipeline to all 1-D spectra.  We use the publicly available rvsao
IRAF external package version 2.7.8 \citep{kur98}.  We run the cross-correlation tool
{\tt xcsao} on our simulated spectra.  The algorithm performs a Fourier
cross-correlation between the ``observed'' (simulated) spectra and a
user-defined library of template spectra.  We obtain the template library used
in the cross-correlation from the simulation itself.  For the first pass, we
pick 6 templates chosen to mimic the 6 galaxy templates used in the
cross-correlation analysis of the SDSS spectroscopic
pipeline\footnote{Templates 23 to 28 in the website: {\url {http://www.sdss.org/dr7/algorithms/spectemplates/index.html}} }.  Using
templates from the simulation instead of the original SDSS templates improved
the number of correct redshifts by $10\%$.  The limitation of the SDSS
template basis is that it was chosen for low redshift spectroscopy, and is not
sufficient for redshifts greater than 1 or so.  In the second pass, we added
three templates from the simulations picked as the brightest templates above
redshift 1.4 for which the cross-correlation coefficient --- the R statistic
described below --- was less than 2.5.  The additional templates doubled the
number of correct redshifts above 1.4.

The cross-correlation analysis can be refined around certain wavelengths by
giving it an initial redshift guess (by setting the parameter {\tt czguess})
to start the search.  We perform the analysis five times with: no guess, {\tt
  czguess} = 0.4, {\tt czguess} = 0.8, {\tt czguess} = 1.2 and {\tt
  czguess}=1.6.  We then choose which redshift estimate to keep based on the
value of the R statistic, output by the pipeline.  The R statistic, introduced
by \citet{ton79} (cf. Eq. 23 of that work), is a measure of the strength of
the cross-correlation given by the ratio of the height of the assumed true
peak in the correlation to the average height of spurious peaks.  R varies
from 1 to several hundred in our simulation, and as we show later, $R > 6$
corresponds to $>99\%$ correct redshifts.

We have performed our analysis for a number of settings of the spectroscopic
pipeline, but only show results for three main cases, defined as follows:
\begin{itemize}
\item {\it Fiducial Pipeline:} $\zspec$s estimated using the 6+3=9 templates and 
the five redshift guesses described above. Yields the highest completeness
for $z>1.4$.
\item {\it Comb2 Pipeline:} $\zspec$s estimated using the 6+3=9 templates and
only running {\tt xcsao} twice, with {\tt czguess = 0.4} and {\tt czguess = 0.8}.
Yields the highest overall completeness, but the lowest completeness at low
and high redshift.
\item {\it Original Pipeline:} $\zspec$s estimated using the 6 
original templates and only four redshift guesses:  
{\tt czguess} = none, 0.4, 0.8 and 1.2.
\end{itemize}

\subsection{Photometric redshifts} \label{sec:photoz}

There exists a cornucopia of publicly available photometric redshift
estimation algorithms.
For recent reviews and comparison of methods see e.g. \cite{hil10,abd11}.
We consider two different photo-z algorithms that broadly span the
space of possibilities.
We use a training-set fitting method with a very large training set
and basic template-fitting code without any priors, which we briefly
describe below.

\subsubsection{Training-set redshift estimators}\label{sec:train}

The basic setup of training-set based redshift estimators is to use a sample
with known spectroscopic redshifts to estimate the free parameters of
a function relating the observables (in our case the magnitudes of the galaxies)
to the redshifts.
After the best-fit free parameters have been determined, the function can be
applied to the data for which no spectroscopic redshifts are available, known
as the photometric sample.
For this paper, we use an artificial neural network as our training set method,
and we leave the details to Appendix \ref{sec:neunet}.

\subsubsection{Template-fitting redshift estimators}\label{sec:templ}

Template-fitting estimators derive photometric redshift estimates by comparing
the observed colors of galaxies to colors predicted from a library of galaxy
spectral energy distributions.  We use the publicly available   {\it LePhare}
photo-z code\footnote{\url{
  http://www.cfht.hawaii.edu/~arnouts/LEPHARE/lephare.html}} \citep{arn99,ilb06}
as our template-fitting estimator.  We chose the extended {\tt ${\rm CWW_{KINNEY}}$} template
library, which comes with {\it Lephare} because it
yielded the best photo-zs for our simulation.
This library includes the four CWW templates \citep{col80} extended to
the IR and UV using templates from \cite{bru03} and six \cite{kin96}
starburst templates.

We note that a variety of public template-fitting codes are available
\citep[e.g.][]{coe06,fel06}, and each includes many options of 
template libraries, extinction laws, priors, etc. 
For a discussion on propagation of template-fitting uncertainties to redshift
uncertainties see \cite{abr11}.

We emphasize that both categories of photo-z estimators should be
thought of as performing the same function, of applying a model to
describe the data. 
For training set methods, the model is the training sample, whereas
for template-fitting methods, the model is the set of templates and priors utilised.
By construction, the training-set method we use has an excellent
training set, and as a result, performs nearly optimally. 
Conversely, for the template-fitting method, we do not take many pains
to optimize the template library nor to find appropriate priors, and
thus, the template-fitting code performs substantially worse.
With perfect templates and perfect priors, template-fitting should equal the
performance of training set methods with perfect training.
It is not our intention to suggest that any of these estimators is
fundamentally superior, we just wished to test our analysis in 
optimistic and pessimistic regimes.
Most importantly, as we show in later Sections, {\it results are
independent of the photo-z estimators used} despite the significant
differences in performance.

\subsection{Effect on the Cosmological Parameters} \label{sec:wl}

In Appendix \ref{sec:wl_app} we review the formalism from BH10 to calculate
the biases in the observed weak lensing power spectra, and hence in the
cosmological parameters, given some arbitrary source systematic
error. 
Here, we are of course interested in the systematics due to imperfect spectroscopic redshifts. 
It is beyond the scope of the paper to consider the
  biases for other cosmological probes, but we believe that the WL
  results will provide a useful baseline for what to expect from a
  single probe. In addition, we expect that a joint analysis might allow for self-calibration of the biases.

The fiducial weak lensing survey corresponds to expectations from the Dark
Energy Survey, and assumes 5000 square degrees (corresponding to $\fsky\simeq
0.12$) with tomographic measurements in $B=30$ uniformly wide redshift bins
extending out to $z_{\rm max}=2.0$.  The effective source galaxy density is 12
galaxies per square arcminute, while the maximum multipole considered in the
convergence power spectrum is $\ell_{\rm max}=1500$. The radial distribution
of galaxies, required to determine tomographic normalized number densities
$n_i$ in Eq.~(\ref{eq:C_obs}), is determined from the simulations and shown in
Fig.~\ref{fig:nzq}.

We consider a standard set of six cosmological parameters with the following
fiducial values: matter density relative to critical $\Omega_M=0.25$, equation
of state parameter $w=-1$, physical baryon fraction $\Omega_B h^2=0.023$,
physical matter fraction $\Omega_M h^2=0.1225$ (corresponding to the scaled
Hubble constant $h\equiv H_0/(100\kmsMpc)=0.7$), spectral index $n=0.96$, and
amplitude of the matter power spectrum $\ln A$ where $A=2.3\times 10^{-9}$
(corresponding to $\sigma_8=0.8$).

As detailed in Appendix \ref{sec:wl_app}, we add the (unbiased)
Planck forecasted constraints on the cosmological parameters to those of the
DES. The fiducial (combined) constraint on the equation of state of dark energy
assuming perfect knowledge of photometric redshifts is $\sigma(w)=0.055$.

\section{Spectroscopic success and failure}\label{sec:results1}

In this section we define the basic concepts regarding
  successful and failed galaxy spectra, and the accompanying rates.

\subsection{Spectroscopic success rate}\label{sec:resspec}

The spectroscopic analysis for the fiducial simulation parameters (16200 secs
integration; 9 templates; no manual correction of spectra) yields about
$74\%$ correct spectroscopic redshifts (defined as redshifts for which
 $|\zspec-\ztrue| < 0.01$). 
In a real survey, one can only choose redshifts based on some quality 
flag, which is the cross-correlation R statistic (described in 
Sec. \ref{sec:iraf})  in our case. 
We thus define two success metrics:
\begin{itemize}

\item {\it True spectroscopic success rate ($\ssrt$)}: the fraction of galaxies 
with correct redshifts.

\item {\it Observed SSR ($\ssro$)}: the fraction of galaxies with R greater than 
a certain value.
Unless stated otherwise, we set the value to 6.0.

\end{itemize}

In Fig. \ref{fig:ssr}, we show the true SSR as a function of true redshift
(left panel), observed i-band magnitude (center panel) and cross-correlation
strength (right panel).  The left panel shows that the $\ssrt$ generally worsens
with higher redshift, and the 'hiccups' in the curves are directly caused by
different spectral lines which enter and leave the observed spectral range, as
discussed in Sec.~\ref{sec:specintro}.  The central panel shows the expected
result that the spectroscopic success rate plunges beyond certain depth.
Finally, the right panel shows that the true SSR increases monotonically with
cross-correlation statistic R, showing that we can use R to select an accurate
redshift sample with high confidence.

\begin{figure}
\includegraphics[scale=0.45,angle=-90]{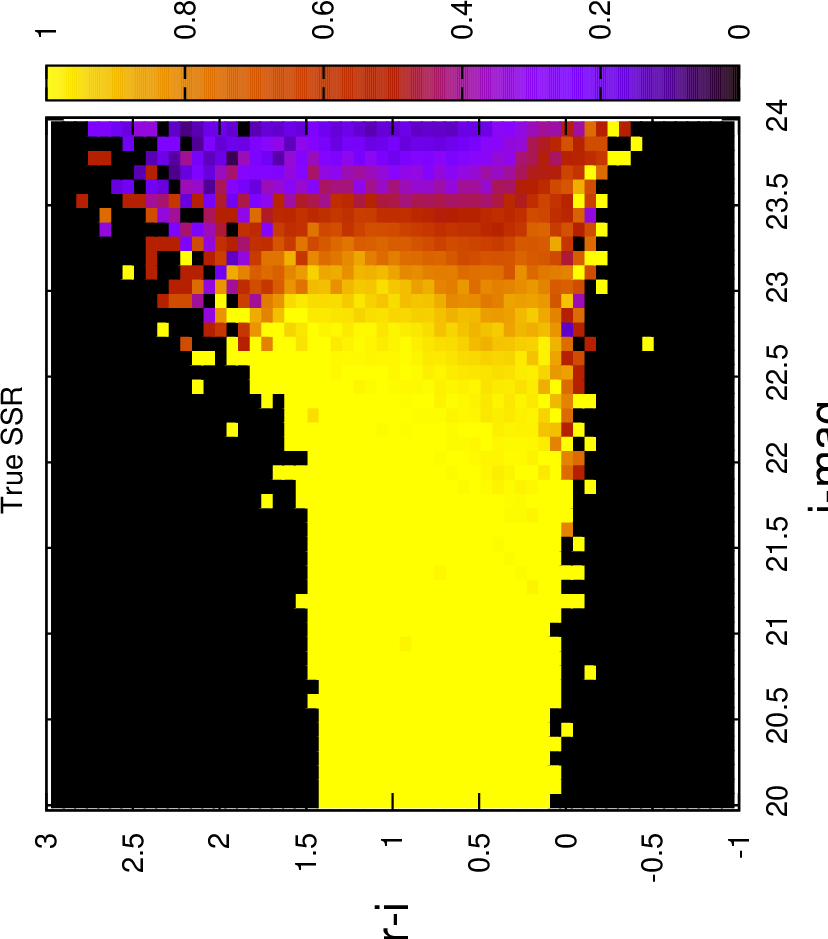}\vspace{0.99cm}
\includegraphics[scale=0.45,angle=-90]{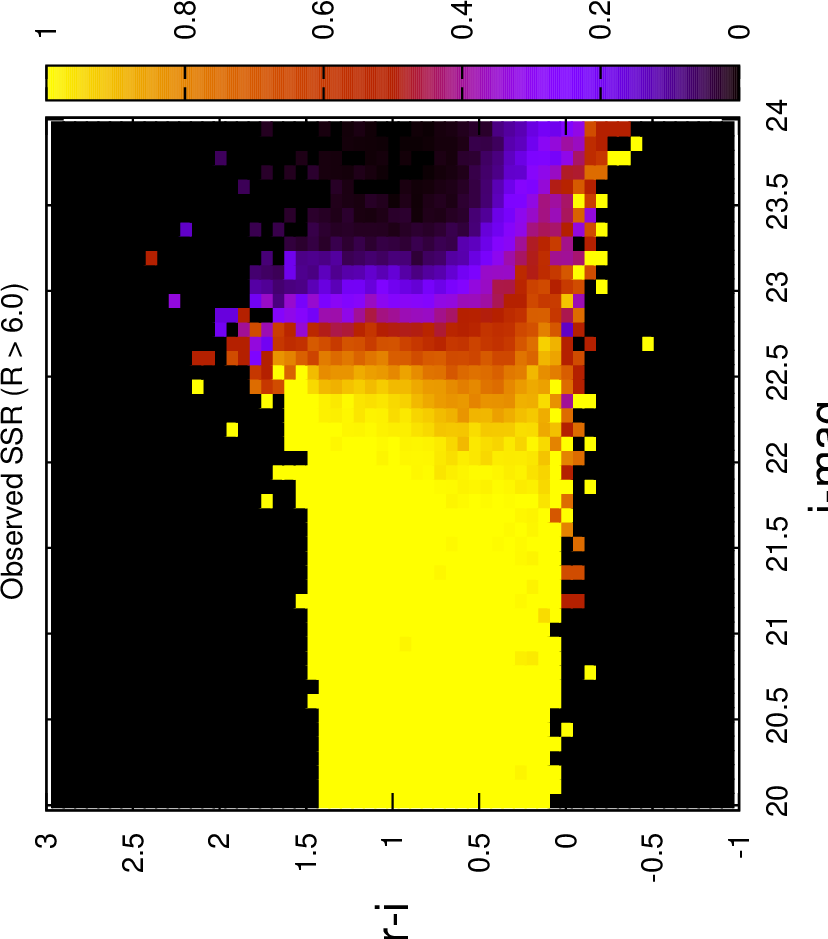}\vspace{0.99cm}
\caption{Top panel: True spectroscopic success rate ($\ssrt$), defined as fraction 
of correct redshifts as a function of true redshift. Bottom panel: Observed 
SSR ($\ssro$), defined as fraction of galaxies with correlation $R> 6.0$. 
Both results assume the Fiducial pipeline settings (cf. Sec. \ref{sec:iraf}) of 
16200 secs of integration time with the 3 additional 
templates. } \label{fig:ssrcol}
\end{figure}

In Fig.~\ref{fig:ssrcol} we show the true and observed SSRs as a function of
$i$-magnitude and $r$--$i$ color.  The top panel
shows that virtually all the incorrect redshifts are at the faint end of 
the color-magnitude diagram, with slight color dependence. 
In particular, at the bluest end ($r$--$i\sim 0$) we see a region of low 
SSR extending to $i\sim 22$. 
This is typically caused by the lack of an appropriate template to describe
certain galaxy populations.%

The {\it observed} SSR, shown in the bottom panel of Fig.~\ref{fig:ssrcol}, shows a
more pronounced color variation.  We can see that the bluer colors,
corresponding to late spectral types, which have significant emission
features, yields highest $\ssro$.  Conversely, the redder colors have the lowest
$\ssro$.  
As mentioned previously, early type galaxies have virtually no emission
lines, and hence are identified by absorption features.  Intermediate types
can have weak emission lines, but usually have weaker absorption features as
well, which makes it difficult to determine a spectroscopic redshift for them.

Because of our stringent choice of cut, the sample with $R>6.0$ contains a
fraction 0.53 of the total galaxies and has $99.6\%$ correct spectroscopic
redshifts.  For comparison, if we define samples by the cuts $R>5.0$ and
$R>4.0$ these would contain a fraction of 0.60 and 0.73 of total galaxies with
$98.6\%$ and $93.2\%$ correct redshifts, respectively.
Faint, intermediate-type galaxy spectra yield the majority of the incorrect 
redshifts that escape the R selection.

\begin{figure}
\includegraphics[scale=0.35,angle=-90]{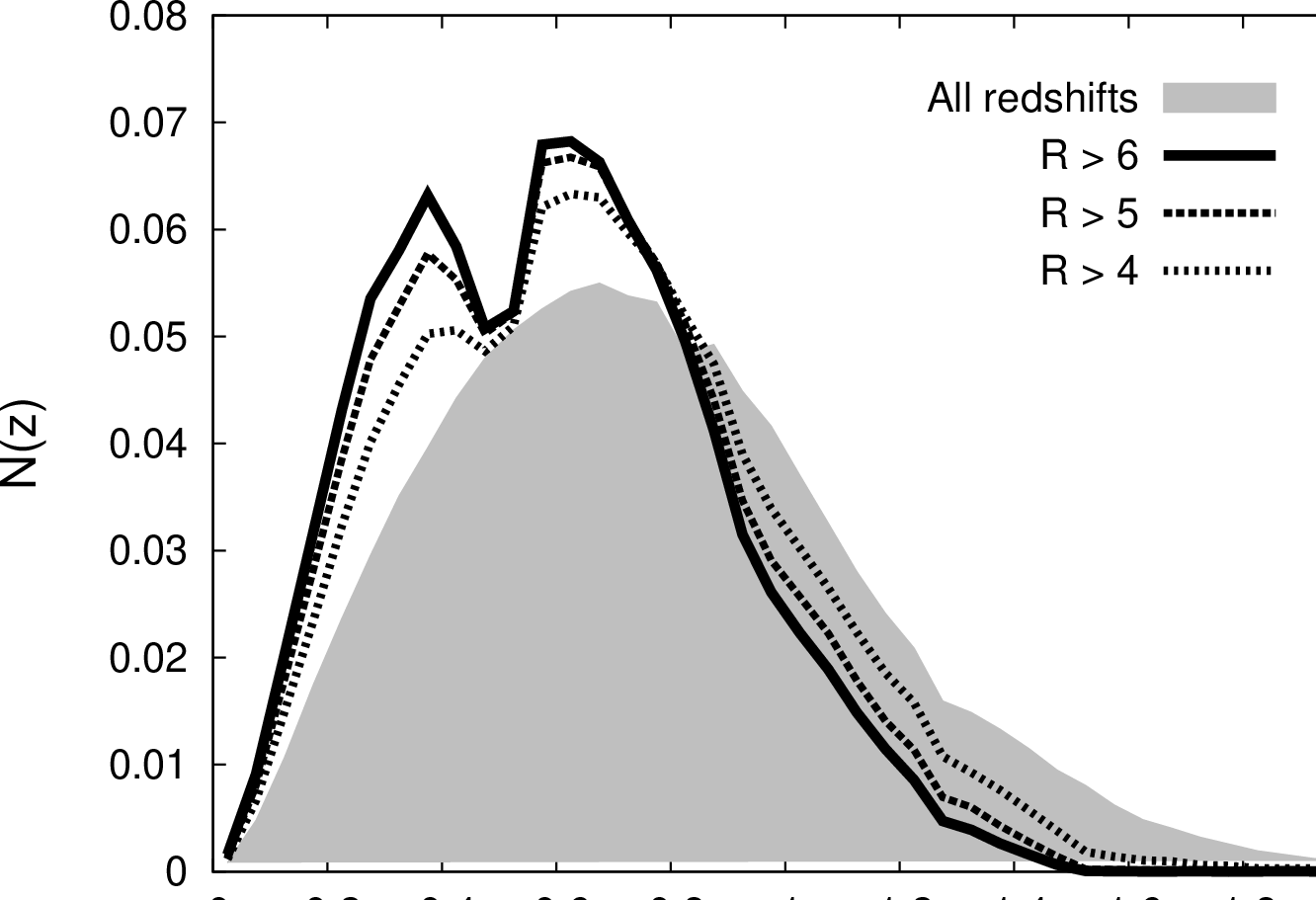}\vspace{0.8cm}
\includegraphics[scale=0.35,angle=-90]{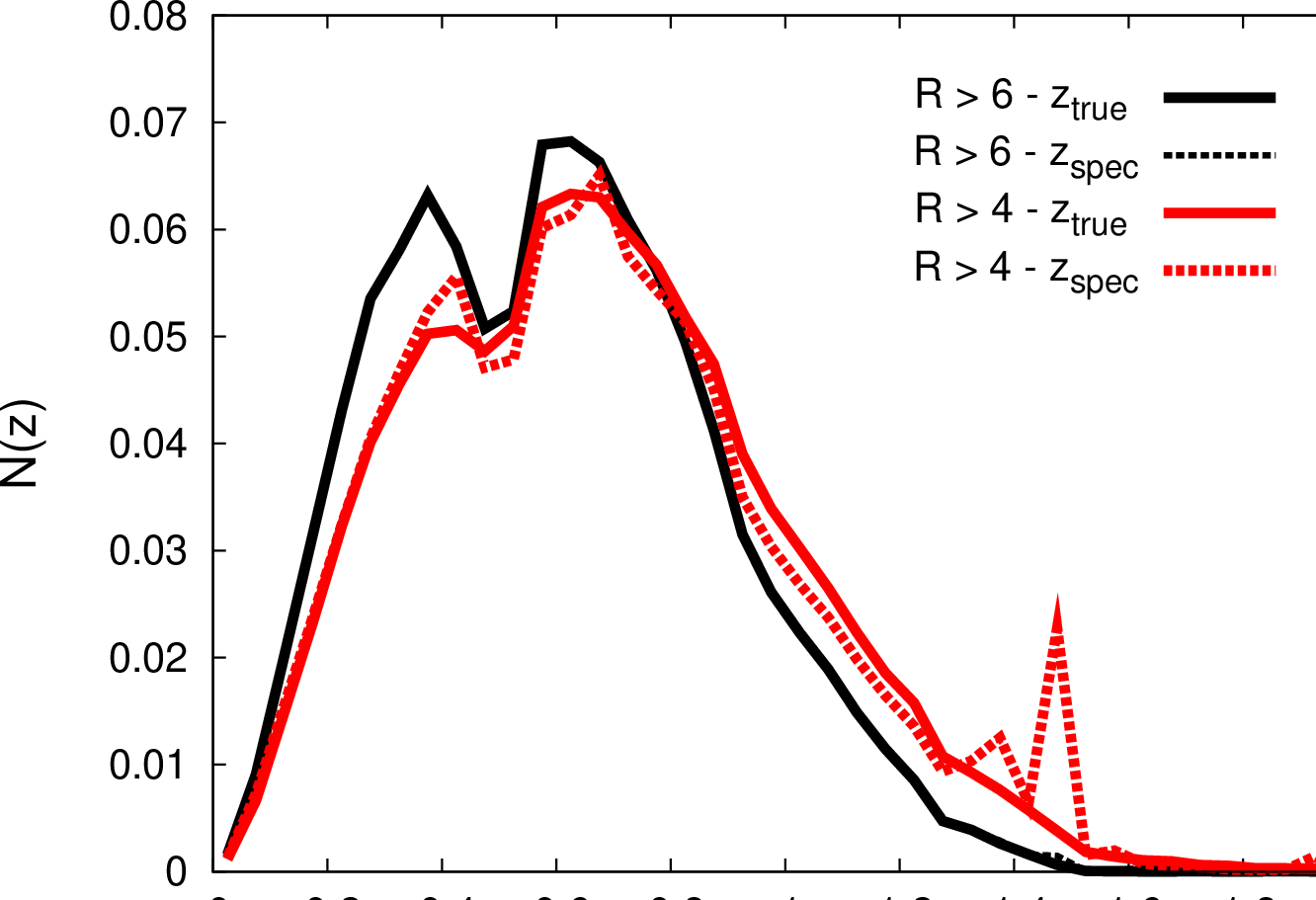}\vspace{0.7cm}
\caption{Top panel: Distributions of true redshift for 
all galaxies (shaded area), galaxies with $R>6$ (solid line), galaxies with $R > 5$ 
(dashed line) and galaxies with $R>4$ (dotted line). 
Bottom panel: Distribution of true redshift (solid lines) and spectroscopic redshift 
(dashed lines) for the $R>6$ sample (black) and the $R>4$ sample (red - gray).
} \label{fig:nzq}
\end{figure}

In the top panel of Fig. \ref{fig:nzq} we show the effect of applying quality
cuts based on the statistic R to the true redshift distribution.  More
stringent (higher R) cuts preferentially remove galaxies from regions where
less significant spectroscopic features fall inside the spectrograph window
(as explained in Sec. \ref{sec:specintro}).  The bottom panel shows that the
less stringent cuts allow for a higher fraction of incorrect redshifts, which 
have a visible impact in the redshift distribution even though $93.2\%$ of the
redshifts are correct.

\subsection{Where do the wrong redshifts go?}

\begin{figure*}
\includegraphics[scale=0.35,angle=-90]{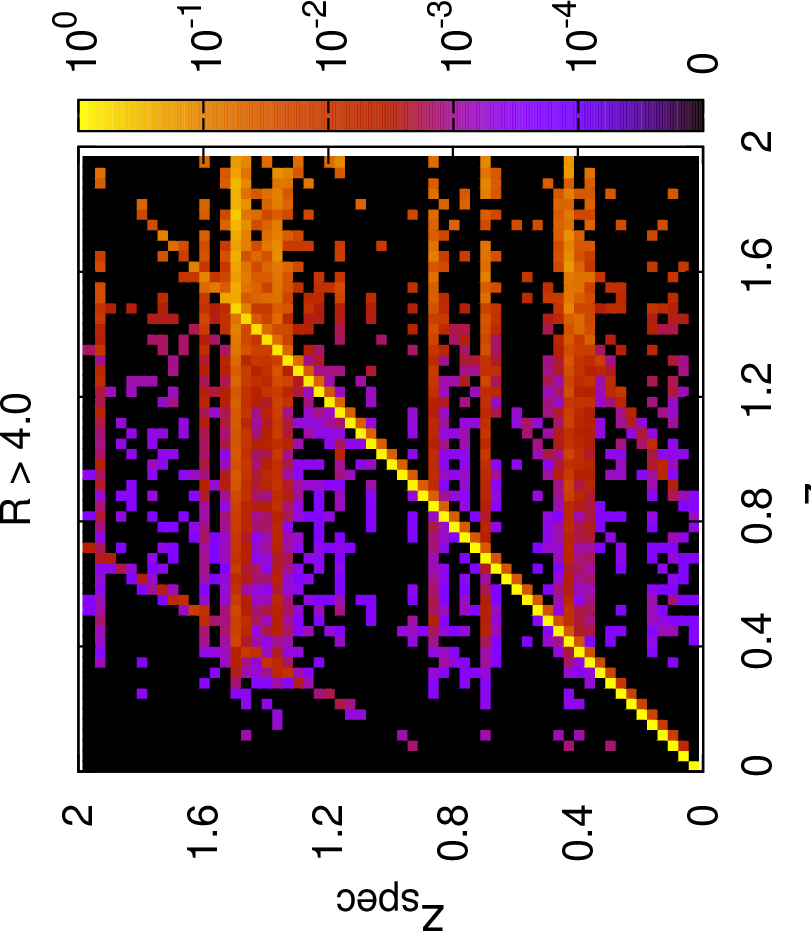}\hspace{0.3cm}
\includegraphics[scale=0.35,angle=-90]{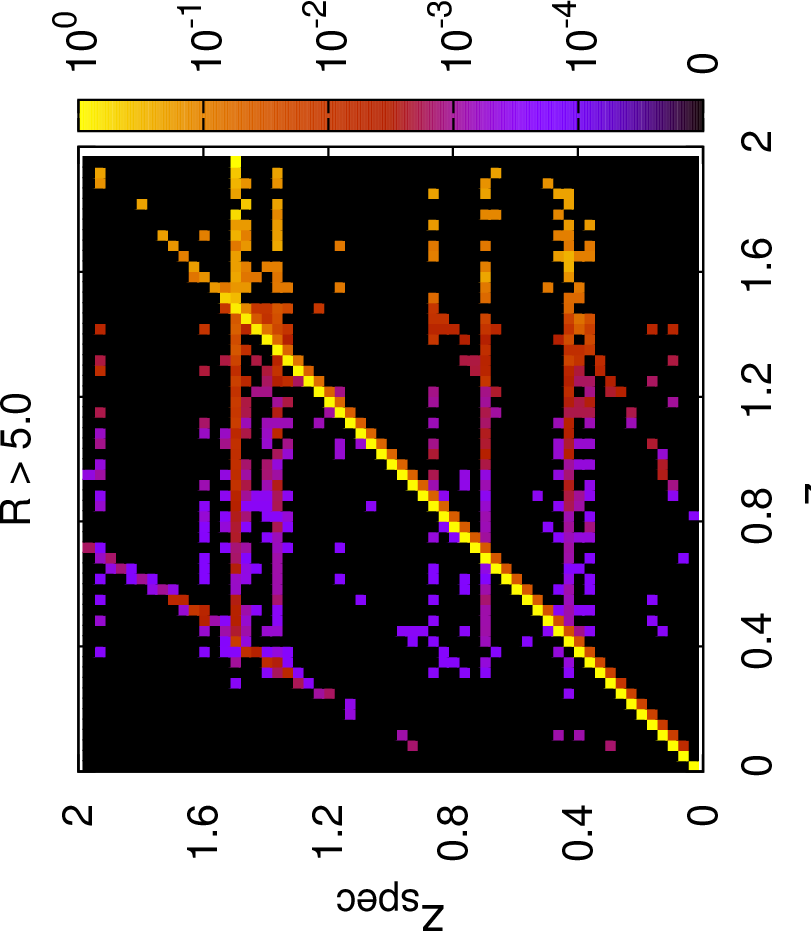}\hspace{0.3cm}
\includegraphics[scale=0.35,angle=-90]{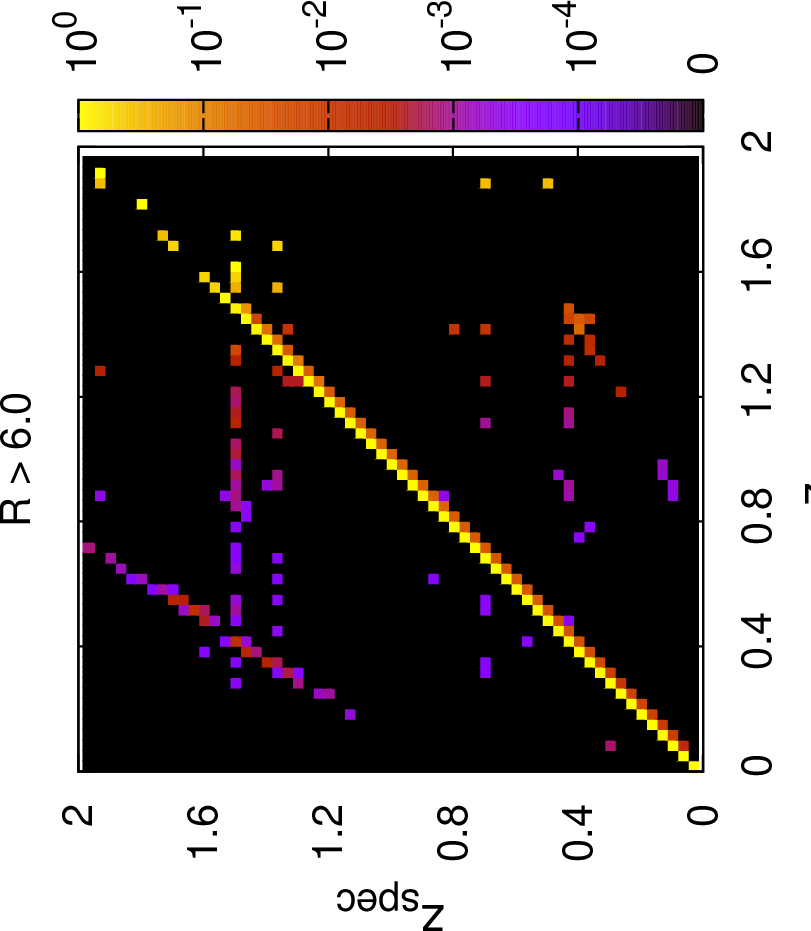}\vspace{0.5cm}
\caption{Leakage matrices ($\pzst$) for the training sets selected by the cuts
  $R>4.0$ (left panel), $R>5.0$ (center panel), and $R>6.0$ (right panel). The
  spectroscopic redshifts were calculated using 16,200 secs exposures with the
  full set of 9 templates in the spectroscopic pipeline, corresponding
    to our Fiducial pipeline.  } \label{fig:zzp}
\end{figure*}

We show the spectroscopic leakage matrices ($\pzst$) for several cuts in the 
R statistic for our Fiducial pipeline scenario in Fig.\ \ref{fig:zzp}. 
The spectroscopic redshift errors, which correspond to any 
departures from the $\zspec=\ztrue$ (diagonal) line, clearly make 
interesting and definite patterns:

\begin{itemize}
\item {\it Atmospheric line confusion:} Horizontal features in
  Fig.\ \ref{fig:zzp}, when many different values of $\ztrue$ are
  misinterpreted as a single $\zspec$, correspond to cases where 
  residuals from subtraction of atmospheric lines are
  confused with actual features in the galaxy spectrum.
\item {\it Galaxy line misidentification:} Diagonal lines in
  Fig.\ \ref{fig:zzp} (excepting the $\zspec=\ztrue$ diagonal, of course)
  correspond to the cases where the pipeline misidentifies lines of the galaxy
  itself due to limited spectroscopic coverage and S/N
  (cf.\ Sec.\ \ref{sec:lines}).  For example, the diagonal trend from
  $(\ztrue, \zspec)=(0, 0.8)$ to about $(0.7, 2.0)$ corresponds to the
  pipeline classifying H$\alpha$ emission lines as [OII] lines.  A corresponding
  feature due to [OII] being incorrectly classified as H$\alpha$ can be seen
  starting at (0.8, 0) in the plots.  Galaxy line misidentification seems to
  be a much smaller issue than atmospheric line confusion for our simulation.

\end{itemize}

\begin{figure}
\includegraphics[scale=0.55,angle=-90]{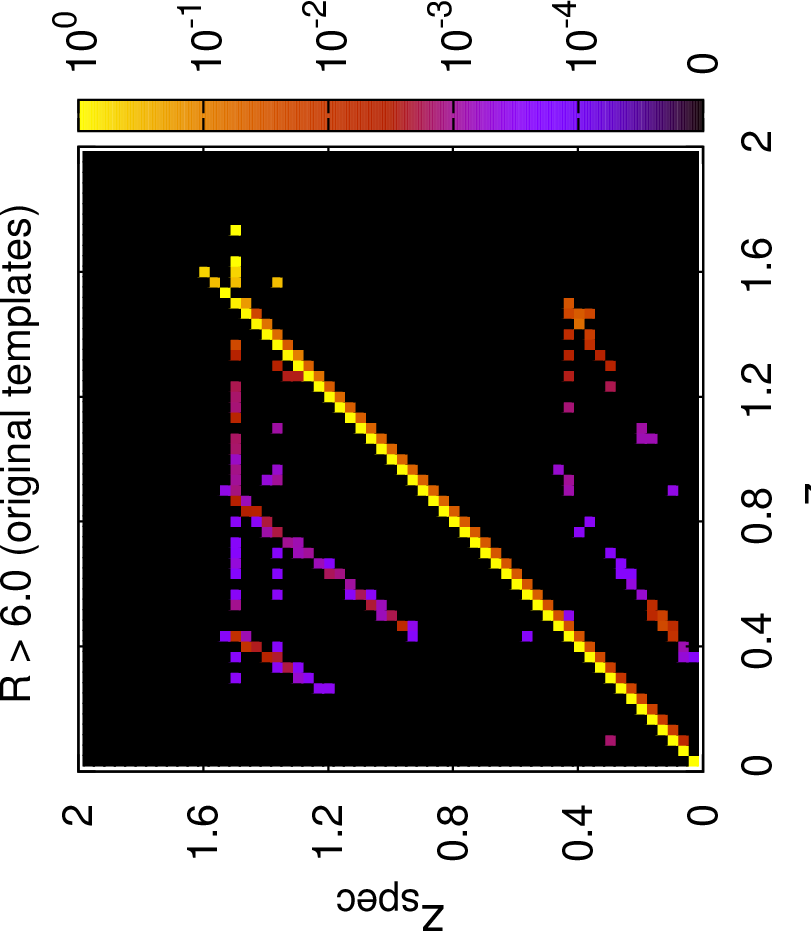}\vspace{0.5cm}
\caption{Same as Fig. \ref{fig:zzp} except for the Original pipeline, where 
  only the 6 original templates were used, and only 4 different values 
  of {\tt czguess} (no guess, 0.4, 0.8 and 1.2) were used in the {\tt rvsao} run. 
  Without the 3 additional templates, no strong correlations were found for 
  $\zspec >1.5$, which, in particular, implied that no galaxies were incorrectly assigned
  $\zspec >1.5$.  } \label{fig:zzpfid}
\end{figure}

The exact distribution of the wrong redshifts depends on the noise levels
assumed and details of the spectroscopic analysis.  As described in Appendix
\ref{sec:pipe}, we assumed a constant mean atmospheric emission and
absorption, but in reality the observing conditions vary.  The distribution of
wrong redshifts also depends on details of the spectroscopic analysis.  In
Fig. \ref{fig:zzpfid} we show the $\pzst$ matrix for the Original pipeline,
described in Sec. \ref{sec:iraf}, which only uses the original 6 spectral
templates (but not the 3 templates added to increase completeness for
$z>1.4$.)  In addition, it does not use the {\tt czguess=1.6} results, which
have the effect of increasing the probability that the pipeline will assign a
high redshift to a galaxy.  The Original pipeline is not optimized in any way
towards high-z completeness, and as a result it finds no spectroscopic
redshifts above $z=1.6$.  Conversely, the Fiducial pipeline (cf. right plot in
Fig. \ref{fig:zzp}), does find some redshifts above $z=1.6$, but at the cost
of increasing the number of objects being incorrectly assigned very high
values of spectroscopic redshifts and the number of objects at high redshifts
being assigned very low redshifts.  As we discuss in Sec. \ref{sec:wlres}, the
Original pipeline yields a bias in $w$ a factor of two smaller than the
Fiducial pipeline.

There are two points to take from this section.  First, wrong spectroscopic
redshifts occupy preferred regions of the ($\ztrue$, $\zspec$) plane.  Since
the exact redshift error distribution depends on the details of the
spectroscopic analysis and observing conditions, it is challenging to
accurately predict the spectroscopic redshift errors in real surveys.  Hence,
our conclusions concerning the impact of wrong redshift are necessarily only
rough estimates.  Second, increasing the completeness at high redshift can
come at the expense of introducing more catastrophic spectroscopic redshifts.
As we shall show in Sec. \ref{sec:wlres}, this is a very high price to pay,
and can severely increase biases in cosmological parameter constraints.

\section{Spectroscopic selection matching}\label{sec:results2}

As can be inferred from the top panel in Fig. \ref{fig:nzq}, 
spectroscopic failures alter the redshift distribution 
of the training set significantly, so that one cannot use such a sample 
to estimate the error distributions of the photometric sample directly.
We test two different approaches to correct for the selection effects 
in the training set.

One approach, presented in Sec. \ref{sec:sel} is to cull the photometric sample to remove all galaxies 
that are not represented in the training set (the set of high-confidence spectroscopic redshift galaxies).
The other approach, described in Sec. \ref{sec:weights} is to apply
weights to the spectroscopic sample so that the statistical properties
of the weighted spectroscopic galaxies closely match those of the
photometric sample. 
Thus, in the first approach, the photometric sample is modified to
match the spectroscopic sample, whereas in the second approach, it is
the spectroscopic sample that is modified.

\subsection{Culling approach to selection matching}\label{sec:sel}

In the culling approach, we use a neural network (described in Appendix \ref{sec:neunet}) to accomplish 
the selection matching.
What we want is to be able to classify galaxies in the photometric sample 
in the same way they were classified in the training set, that is, we need 
to estimate the cross-correlation strength R statistic for them.

To be more realistic, instead of using R, we map the R values 
into a new quality parameter Q. 
The Q parameter is {\it discrete}, and roughly matches the more standard 
quality flags of real surveys (e.g. VVDS, DEEP2). 
It also has the advantage of having a more limited range than 
the R statistic, which has no upper limit.
The mapping we use is as follows:

\begin{eqnarray*}
R \geq 6 &\Longleftrightarrow & Q=4\\
5<R<6    &\Longleftrightarrow & Q=3\\
4<R<5    &\Longleftrightarrow & Q=2\\
3<R<4    &\Longleftrightarrow & Q=1\\
0<R<3    &\Longleftrightarrow & Q=0
\end{eqnarray*}

\noindent The relationship between $\ssrt$ and Q classification depends
slightly on the exposure time, but can be inferred from the right plot
of Fig. \ref{fig:ssr}. 

Following standard neural network procedure, we split the spectroscopic sample
into two parts (of equal size), the training and validation samples.  
As described in Appendix \ref{sec:neunet}, we use the $griz$ magnitudes 
as the inputs for the neural network, which then outputs an estimate for $Q$.
For simplicity, we only perform a single neural net run, though
the average of multiple runs is expected to yield best results.

After the neural net run converges, we apply the best-fit function to the
complete spectroscopic sample and to the photometric sample to obtain
estimates of the Q coefficient, hereafter ${{\rm Q}_{\rm est}}$, for all the
galaxies. 
Note that ${{\rm Q}_{\rm est}}$ is continuous, whereas Q is discrete.

The matching between Q and $\qest$ is very good for the ${\rm Q}=4$
galaxies. 
For the 16200 sec exposures, the $\sse$ width of the ${\rm Q}- \qest$
distribution is 0.05, and only $15\%$ of the galaxies with ${\rm Q}=4$ have 
$|{\rm Q}- \qest|>0.5$.
The matching is less accurate for the galaxies for the lower Q, with the
worst case being the ${\rm Q}=3$ sample, for which $\sse=0.88$.
Interestingly, the redshift success statistics for the Q-selected
sample are very similar to those of a $\qest$ selected sample.
Most importantly, the spectroscopic success rate increases
monotonically with $\qest$ (not shown), making it an accurate classifier of
redshift confidence. We leave more detailed comparisons between Q and
$\qest$ for future work.


We apply cuts on $\qest=1.5$, 2.5, and 3.5 to both spectroscopic and
photometric samples.  With the 16200 sec exposures, the corresponding True
SSR for the galaxy samples is 0.996, 0.978 and 0.914, respectively, with a 
corresponding fraction of objects relative to the total of 0.463, 0.586 and 
0.751 in the three cases.  
For the 48600 sec exposures, we find True SSRs of 0.996, 0,978 and 0.936 
respectively, with corresponding fractions of objects retained of 0.655, 
0.808 and 0.960.

The next step is to investigate the impact of the selection to the weak
lensing analysis. 
We break up the process into several parts, for clarity:
\begin{itemize}
\item If a training set based method is to be used for calculating photo-zs,
  the first step is to use the training sample with the desired $\qest$ cut to
  derive photometric redshifts for the matched photometric sample
  (cf. Sec. \ref{sec:pztrain}).  This step may be skipped if a pure
  template-based algorithm is being used.

\item Next, we calculate the WL constraints for the photometric sample
  selected with the $\qest$ cut and compare that to what we get for the full
  sample.  Constraints degrade both from the reduction in the total number of
  objects as well as with the shift of the redshift distribution towards lower
  redshifts (cf. Sec. \ref{sec:wlres}). 

\item The next step is to assess the bias resulting from differences in the 
selection of the spectroscopic and photometric samples as well as the 
biases due to wrong redshifts. (cf. Sec. \ref{sec:wlres}).
\end{itemize}

\subsubsection{Photo-z training}\label{sec:pztrain}

We use a neural network photo-z estimator to exemplify the impact of selection
matching and wrong redshifts on training-set based photo-z estimation
(cf. Sec. \ref{sec:train}).  For simplicity, we assume that the photo-zs for
the photometric sample should only be calculated for the subset of galaxies
surviving the selection cuts of the previous section.  In other words, we
require that the spectroscopic training sample and the photometric sample have
matching selections.  We thus define three sets of spectroscopic and
photometric samples, specified by the spectroscopic quality cuts on $\qest$ of
$\qest >$ 3.5, 2.5, or 1.5.

To separate the effects of selection matching from the effect of wrong
redshifts, we estimate the photo-zs twice.  First, we assume we have the true
redshifts for all galaxies passing the $\qest$ cuts, to isolate potential
biases due to the spectroscopic selection matching.  Then, we perform the
photo-z training on the actual spectroscopic redshifts, to gauge the
additional impact of wrong redshifts.

Table \ref{tab:ztrain} shows the $1\sigma$ photo-z scatter for the samples
defined by the $\qest$ cuts.  The two $\ztrue$ columns correspond to
the scenarios where the true redshifts were used in the training.  
The scatter is defined as the dispersion in the distribution of 
$(\ztrue-\zphot)$ for both the training sample and photometric sample.  
As expected, the photo-z scatter of the training sample is in
excellent agreement with the scatter of the photometric sample, suggesting
that both samples have close to identical photo-z properties and that the
selection matching does not introduce any biases.  
Furthermore, the scatter improves as we apply more stringent cuts on $\qest$.  
The decrease in scatter is as expected, since the objects with low $\qest$ 
are typically the faintest.

The three $\zspec$ columns in Table \ref{tab:ztrain} show the more realistic
case where the actual spectroscopic redshifts (wrong redshifts included) was
used to train the photo-zs.  In the $\zspec({\rm Train})$ we show the scatter
in the training set calculated as the dispersion in the $(\ztrue-\zphot)$
distribution, which we can see is in excellent agreement with the scatter of
the photometric sample shown in the fifth column.  Comparing the dispersion of
the $\zspec({\rm Photo})$ and $\ztrue({\rm Photo})$ cases, we see that the
presence of wrong redshifts degrades the photo-zs of the photometric sample by
as much as $20\%$ in the case of the $\qest>1.5$ cut.  The degradation is
reduced for the more stringent cuts as the fraction of wrong redshifts is
reduced.

In reality, one does not know the true redshifts for the training set,
but only the spectroscopic redshifts.
Hence, the scatter in the training set photo-zs would be estimated
using the spectroscopic redshifts, as the dispersion in the $(\zspec-\zphot)$
distribution.
We show this estimate of the scatter in the $\zspec({\rm Train^*})$ column.
We see substantially larger values of the scatter compared to the 
$\zspec({\rm Photo})$ column, for all $\qest$ cuts.  
The point is that the neural network is not substantially affected by
the wrong redshifts, so that the true scatter does not degrade.
However, our estimate of the scatter using the spectroscopic redshifts
is strongly affected, as the wrong spectroscopic redshifts show up
as catastrophically incorrect redshifts, which we can often remove.

We therefore conclude that the use of training samples is
  still justified in the presence of incorrect redshifts. For the sake of
  comparison, the template-fitting photo-zs without any priors have rms
  scatter of 0.229, 0.215 and 0.203 for the samples selected with $\qest
  >1.5$, 2.5 and 3.5, respectively. If the scatter were estimated using the
  spectroscopic redshifts, then the rms would be 0.313, 0,246 and 0.211 for
  the same cases.  A more careful analysis of the template-fitting photo-zs
  could certainly produce better results.  However, as we show in the next
  Section, the cosmological biases are very similar whether template or neural
  net photo-zs are used, and we leave more detailed analyses for future
  work. For more comparisons between these two methods using a catalog with
  similar photometry, see \citet{cun12}.

\begin{table}
\begin{center}
\begin{tabular}{cc|cc|ccc}
\hline\hline \multicolumn{6}{c}{\rule[-2mm]{0mm}{6mm} Photo-z scatter and training set size }\\
\hline\hline \multicolumn{1}{c}{\rule[-2mm]{0mm}{6mm} }   &\multicolumn{2}{c}{\rule[-2mm]{0mm}{6mm} $\ztrue$} &
\multicolumn{3}{c}{\rule[-2mm]{0mm}{6mm} $\zspec$}\\
\hline   Selection  & \rule[-2mm]{0mm}{6mm} Train & \rule[-2mm]{0mm}{6mm} Photo 
& \rule[-2mm]{0mm}{6mm} Train    & \rule[-2mm]{0mm}{6mm} Photo 
& \rule[-2mm]{0mm}{6mm}  Train* \\\hline
\rule[-2mm]{0mm}{6mm}$\qest > 1.5$&0.121  &0.121  &0.149 &0.149 &0.214  \\
\rule[-2mm]{0mm}{6mm}$\qest > 2.5$&0.098  &0.099  &0.105 &0.106 &0.142  \\
\rule[-2mm]{0mm}{6mm}$\qest > 3.5$&0.082  &0.083  &0.081 &0.082 &0.098  \\\hline\hline
\end{tabular}
\caption{Rms scatter of neural network photo-zs for the samples selected by the cuts
  on estimated $\zspec$ quality, $\qest > 1.5$, 2.5, and 3.5. 
    Note that the scatter for the Train*/$\zspec$ column is defined
  as the dispersion in the $\zspec-\zphot$ distribution, whereas it's defined
  as the dispersion in the $\ztrue-\zphot$ for the other
  columns. }
\label{tab:ztrain}
\end{center}
\end{table}

\subsubsection{WL constraints and biases}\label{sec:wlres}

In this section we examine the constraints and biases in the dark energy
equation of state $w$ inferred from weak lensing shear-shear correlations.
The errors in $w$ are caused by our inability to characterize the photometric
redshift error distribution of our sample.  In other words, we must know the
$\pztrue$ error matrix for our photometric sample to high accuracy.  When we
rely on a spectroscopic sample to characterize the error distribution, we are
actually estimating $\pzs$, but this distribution differs from the true error
matrix $\pztrue$ because of issues in spectroscopic selection matching and wrong
spectroscopic redshifts.  We now investigate how these spectroscopic redshift
errors affect the dark energy equation of state measurements.

Table \ref{tab:wstats} shows the $1\sigma$ constraints on $w$ and systematic
errors for several different sample selections.  The results shown used
template-fitting photo-zs described in Sec. \ref{sec:templ}.  For clarity, we
artificially separate the issues due to selection matching from that of the
wrong redshifts as follows: we perform the cosmological parameter forecast
analysis assuming that all redshifts that passed the $\qest$ selection cut
were the correct, true redshifts, thereby explicitly isolating the selection 
matching systematics.  
The results are presented under the $\ztrue$ column in Table \ref{tab:wstats}.
We can see that biases in $w$ are negligible compared to the statistical 
constraints, demonstrating that the neural network can accurately match the 
spectroscopic selection to the photometric sample.  
The table also shows the fraction of galaxies surviving the selection cuts.  
For example, for the 16200 secs exposures, we see that the $\qest>3.5$ cut 
removes more than half of the sample, which results in nearly a factor of 
two degradation in the statistical constraints relative to what is 
achievable with the full sample ($\sigma(w)=0.055$).  
The degradation is so severe because most of the objects removed by the 
cut are at high redshifts.

Next, we examine the impact of wrong redshifts.  As the last column of Table
\ref{tab:wstats} shows, wrong redshifts can be devastating to the weak lensing
constraints.  The bias in $w$ is, perhaps, tolerable only in the $\qest>3.5$
cases.  In the other scenarios one can see that the biases in $w$ are greater
than the $1\sigma$ constraints even with close to $98\%$ correct redshifts
(${\rm SSR}_{\rm T}\simeq 0.98$).

Comparing the 48600 secs and 16200 secs results we see that the magnitude of
the biases in $w$ are set entirely by the spectroscopic success rate
($\ssrt$), regardless of the level of completeness.  This is another reminder
that the emphasis must be on accuracy over completeness.

\begin{table}
\begin{center}
\begin{tabular}{cccccc}
\hline\hline \multicolumn{6}{c}{\rule[-2mm]{0mm}{6mm} Constraints on $w$ (template-fitting photo-zs) }\\
\hline\hline \multicolumn{1}{c}{\rule[-2mm]{0mm}{6mm} {\textbf{ 16200 secs}} }   &\multicolumn{3}{c}{\rule[-2mm]{0mm}{6mm} } &
\multicolumn{2}{c}{\rule[-2mm]{0mm}{6mm} bias$(w)$}\\
\hline   Selection & {\rule[-2mm]{0mm}{6mm} Gal. Frac.}& {\rule[-2mm]{0mm}{6mm} ${\rm SSR}_{\rm T}$ ($\%$)}  & {\rule[-2mm]{0mm}{6mm} $\sigma(w)$} & \rule[-2mm]{0mm}{6mm} $\ztrue$ & \rule[-2mm]{0mm}{6mm} $\zspec$ \\\hline
\rule[-2mm]{0mm}{6mm}$\qest > 1.5$&0.75 & 91.4 &0.07& 0.004 &- 0.52     \\
\rule[-2mm]{0mm}{6mm}$\qest > 2.5$&0.59 & 97.8 &0.09& 0.002 &- 0.13     \\
\rule[-2mm]{0mm}{6mm}$\qest > 3.5$&0.46 & 99.6 &0.10&-0.001 &- 0.02     \\
\hline \multicolumn{1}{c}{\rule[-2mm]{0mm}{6mm} \textbf{ 48600 secs} }   &\multicolumn{2}{c}{\rule[-2mm]{0mm}{6mm} } &\multicolumn{1}{c}{\rule[-2mm]{0mm}{6mm} } &\multicolumn{1}{c}{\rule[-2mm]{0mm}{6mm} } &\multicolumn{1}{c}{\rule[-2mm]{0mm}{6mm} } \\
\hline
\rule[-2mm]{0mm}{6mm}$\qest > 1.5$&0.96&93.6 &0.06  &0.004  &- 0.39     \\
\rule[-2mm]{0mm}{6mm}$\qest > 2.5$&0.81&97.8 &0.07  &0.005  &- 0.15     \\
\rule[-2mm]{0mm}{6mm}$\qest > 3.5$&0.66&99.6 &0.08  &0.003  &- 0.03     \\
\hline\hline
\end{tabular}
\caption{Statistical and systematic errors in the dark energy equation of
  state $w$ for the different $\qest$-selected samples.  The bias results shown used the
  template-fitting photo-zs.  The Gal.\ Frac.\ column indicates the fraction
  of galaxies from the full data set that passed the selection
  cut, and the ${\rm SSR}_{\rm T}$ indicates the fraction of correct redshifts 
  (i.e, fraction for which $|\zspec-\ztrue|<0.01$) in the sample. 
  The true redshifts $\ztrue$ column assumes, artificially, that all galaxies in the 
  spectroscopic sample that passed the $\qest$ cut had perfect spectroscopic redshifts.
  The $\zspec$ column shows the more realistic case where the actual spectroscopic redshifts
  (including the small fraction of wrong redshifts) were used in the calibration of the photo-z
  error distributions.
  Recall that the statistical, marginalized, error in $w$ for perfect
  redshifts is $\sigma(w)=0.055$} 
\label{tab:wstats}
\end{center}
\end{table}

We investigated the dependence of the results on the photo-z estimator by
performing the WL analysis with the neural network photo-zs instead of the
template photo-zs.  The resulting biases in $w$ are shown in the third column of
Table \ref{tab:otherstats}.  Comparing to the fourth column, where we reproduce
the template photo-z biases from Table \ref{tab:wstats}, we see that the
magnitude of the bias is very similar for the two photo-z estimators, despite
noticeable differences in the photo-z error distributions of
 both \citep[see e.g.][]{cun12}.

We also tested the possibility of decreasing the biases by culling photo-z
outliers.  In the presence of wrong spectroscopic redshifts, the culling could
remove not only catastrophic photometric redshifts, but perhaps also identify
the wrong $\zspec$s.  We used the nearest-neighbor error estimator, NNE
\citep{oya08b}, to cull $10\%$ of the sample selected as the galaxies with
largest NNE error, ($\enne$).  Since the fraction of objects to be culled was
fixed, the value of the $\enne$ cut varied for each catalog and photo-z
estimator.  The results are presented in the last two columns of Table
\ref{tab:otherstats}.  For simplicity, we did not
recalculate the fiducial
constraints when deriving the biases for the culled samples; given the
qualitative nature of this analysis, this is a reasonable approximation.  The
NNE cut seems quite effective for the neural network photo-zs, typically
reducing the biases by half. When the NNE culling was applied to the
template-fitting estimator, the effect was negligible for the $\qest >3.5$
case, and relatively small for the other cases, suggesting that the NNE is
only effective for identifying spectroscopic outliers when a training set
based procedure is used.  This is by no means obvious since the NNE is very
efficient at identifying photo-z outliers even when template-fitting methods
are used \citep{oya08b}.  For comparison, we also tested the effect of
applying the same $10\%$ cut using an error estimator from the
template-fitting code itself\footnote{The error estimate we use is the
  difference between the {\tt Z\_BEST68\_HIGH} and {\tt Z\_BEST68\_LOW}
  outputs of the {\it LePhare} code.} - shown in the last column of
Table \ref{tab:otherstats}.  We find that the biases due to wrong
redshifts for the $\qest>$ 1.5, 2.5 and 3.5 cases are reduced to -0.41, -0.086
and -0.014, showing that culling using this error estimator is also
beneficial. 
We emphasize that the improvement from culling outliers is {\it not} due to the
  improvement of the photo-z statistics, but only due to the 
  removal of spectroscopic failures with the culling. The effect on
  the neural net photo-zs was more significant simply because a larger fraction
of failures was removed.
In contrast, note that, in \cite{cun12}, we found that culling
based on photo-z error estimates had little impact on cosmological biases due 
to sample variance in calibration sample, despite the effective identification 
of the photo-z outliers. 

What about removing the wrong redshifts by comparing the photo-zs to
  spectroscopic redshifts? This is an intriguing option, but one must be very
  careful. The outlier islands of the photo-z error distribution often cause
  the largest biases in cosmological parameters \citep{BH10,Hearin}. By
  removing catastrophic objects from the spectroscopic sample, one all but
  ensures that catastrophic islands will not be calibrated properly, thereby
  compromising the cosmological constraints.  Hence, even though the
  comparison to photo-zs can produce a cleaner spectroscopic sample, its
  selection will no longer match that of the photometric sample.  There are
  two ways to salvage the situation. Either one can apply the neural network
  procedure once again to match the selection, or one can directly remove the
  regions of observable space occupied by the objects with the corresponding
  photo-zs. The latter approach is essentially what we have done by performing
  the NNE cuts described in the previous paragraph. Regions of observable
  space for which the difference between $\zspec$ and $\zphot$ are large were
  removed from both the spectroscopic and photometric samples, thereby
  removing the spectroscopic outliers and simultaneously matching the corresponding
  selection in the photometric sample.

Finally, we investigated the dependence of the results on the details of our
spectroscopic pipeline, described in Sec. \ref{sec:iraf}.  We find that our
Fiducial pipeline, despite giving the best high redshift completeness, yielded
the largest biases in $w$, shown in the Table \ref{tab:wstats}.  The different
pipelines yielded consistent trends, and we focus on one particular case, that
highlights the importance of the settings.  The Original pipeline had a
factor of two smaller bias for the $\qest>3.5$ sample.  In the Original
setting, recall that only 6 templates were used.  As can be seen by comparing
the right plot in Fig. \ref{fig:zzp} with Fig. \ref{fig:zzpfid}, the 3
additional templates increased the redshift completeness above $z>1.4$ but
resulted in leakage from the high $\ztrue$ bins to low $\zspec$ bins.  In
particular, some galaxies at $\ztrue \sim 1.9$ were assigned $\zspec$s of $\sim
0.5$ and $\sim 0.7$.  This failure mode was responsible for about 2/3 of the
increase in bias in going from the Original to the Fiducial pipeline.  The
remainder of the difference was due to the fact that the Fiducial pipeline
uses $\tt czguess=1.6$ which has the effect of increasing the probability that
a galaxy will be assigned a high redshift.
As a result, the Fiducial pipeline yields $\zspec$s above 1.5 for several galaxies with $\ztrue <0.8$.
  
We conclude that the commonly adopted approach of maximizing
the completeness is not recommended because it 
leads to the increase of the fraction of wrong redshifts which in turn implies
worse dark energy parameter biases. 


\begin{table}
\begin{center}
\begin{tabular}{ccccccc}
\hline\hline \multicolumn{7}{c}{\rule[-2mm]{0mm}{6mm} Biases in $w$
  (culling catastrophics with NNE ) }\\
\hline\hline \multicolumn{1}{c}{\rule[-2mm]{0mm}{6mm} {\textbf{ 16200 secs}} }& \multicolumn{1}{c}{{} } 
&\multicolumn{2}{c}{\rule[-2mm]{0mm}{6mm} No NNE Cut} & \multicolumn{2}{c}{\rule[-2mm]{0mm}{6mm} NNE Cut}& \multicolumn{1}{c}{\rule[-2mm]{-2mm}{6mm} T. Cut}\\
\hline   {\rule[-2mm]{-5mm}{6mm} Selection} & {\rule[-2mm]{-2mm}{6mm} G. Frac.}& {\rule[-2mm]{-2mm}{6mm} neural}  & {\rule[-2mm]{-2mm}{6mm} templ.} 
& \rule[-2mm]{-2mm}{6mm} neural & \rule[-2mm]{-2mm}{6mm} templ. & \rule[-2mm]{0mm}{6mm} templ.\\\hline
\rule[-2mm]{-5mm}{6mm}$\qest > 1.5$ & 0.75 &- 0.27  &- 0.52 &- 0.19  &- 0.35 &-0.41   \\
\rule[-2mm]{-5mm}{6mm}$\qest > 2.5$ & 0.59 &- 0.13  &- 0.13 &- 0.06  &- 0.11 &-0.09   \\
\rule[-2mm]{-5mm}{6mm}$\qest > 3.5$ & 0.46 &- 0.02  &- 0.02 &- 0.01  &- 0.02 &-0.01  \\
\hline\hline
\end{tabular}
\caption{Biases in the dark energy equation of state $w$ for both the training-set
  and template-fitting  photo-z estimates when the NNE estimator is used to cull outliers
  in $|\zphot-\zspec|$ space. The last column assumes the
  template-fitting photo-zs were culled based on the template-fitting
  error estimates.  
  The 'G. Frac.' column indicates the fraction of galaxies from the full data set
  that passed the selection cut. 
  Recall that the statistical marginalized errors in $w$ for the three $\qest$ 
  cases are 0.07, 0.09 and 0.10 respectively, as shown in Table \ref{tab:wstats}.}
\label{tab:otherstats}
\end{center}
\end{table}

\subsection{Weights approach to selection matching}\label{sec:weights}

In Section \ref{sec:sel}, we matched the selection of the spectroscopic and
photometric samples by culling the photometric sample.  That is, we
selectively removed galaxies from the photometric sample so that it
statistically matched, as closely as possible, the spectroscopic sample.  In
this section we try a more aggressive approach that allows us to keep nearly
the full photometric sample.  Our technique is to weight galaxies in the
spectroscopic sample using the {\tt probwts} method of \cite{lim08} and
\cite{cun09}, so that the statistical properties of these weighted
spectroscopic galaxies match those of the photometric sample.  For convenience
of reference, we briefly describe the {\tt probwts} technique in Appendix
\ref{app:probwts}.

We select a training set by picking galaxies from the spectroscopic sample
with Q above some threshold \qcrit.  We test the reconstruction for several
values of \qcrit.  Following standard {\tt probwts} procedure, we remove the (small)
part of the photometric sample that is determined to have zero overlap with
the spectroscopic sample.  This removes at most a few percent of the
photometric sample, with negligible impact on the statistical constraints.

Note that, in the first approach, with the neural net, all the spectroscopic
sample is used to characterize the spectroscopic selection in observable
space.  The cosmological analysis is then only performed on the sample that
matches the estimated selection.  In the second, we only use reliable spectra,
which we re-weight to match the full photometric sample.  Then, the full
photometric sample is used on the cosmological analysis.  The first approach
is the more conservative one as it throws away photometric data, to keep only
the most reliable sample.  The second approach is more aggressive as it tries
to keep most of the data and only rescale the training set.

\begin{figure}
\includegraphics[scale=0.33,angle=-90]{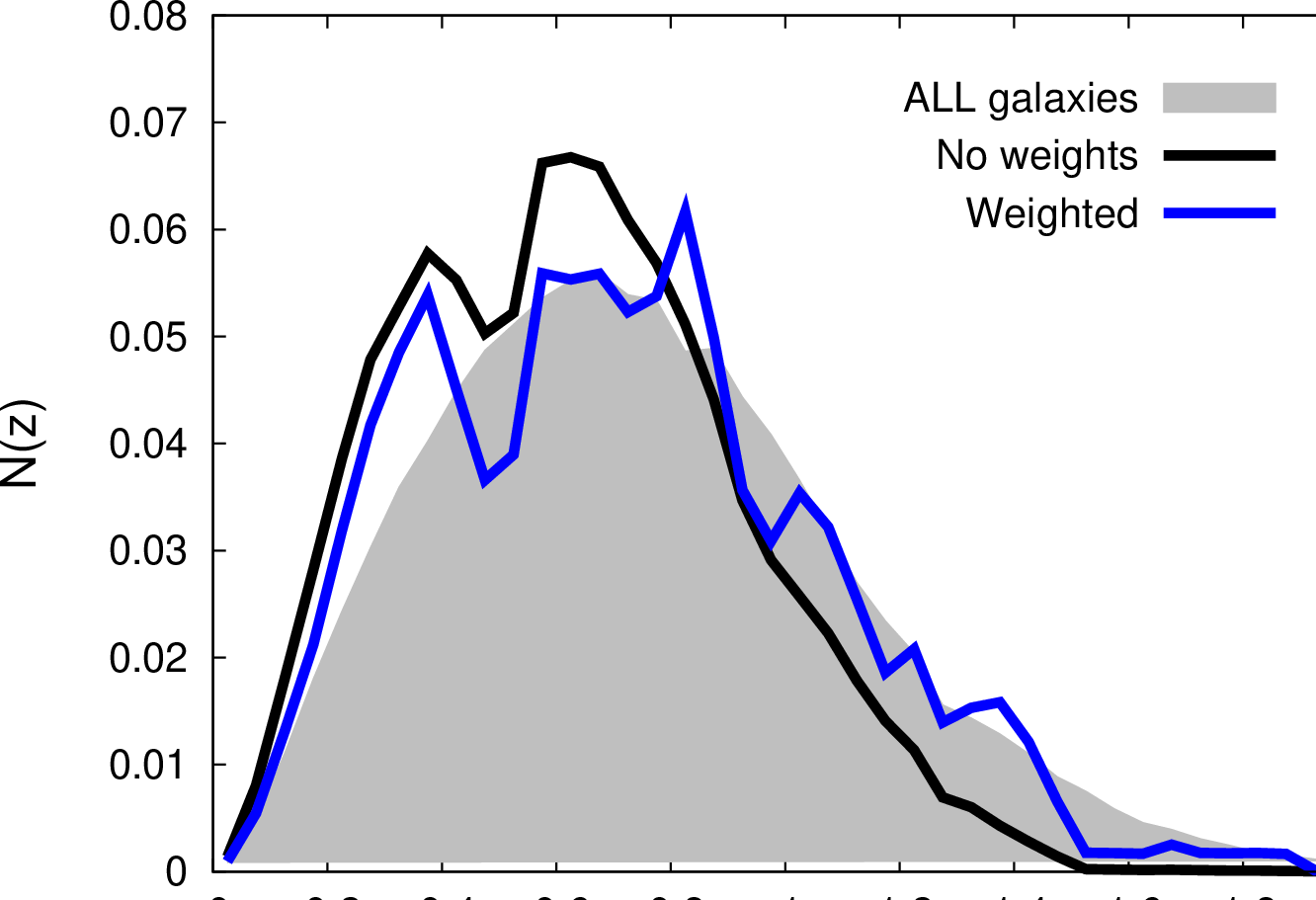}\vspace{0.5cm}
\includegraphics[scale=0.33,angle=-90]{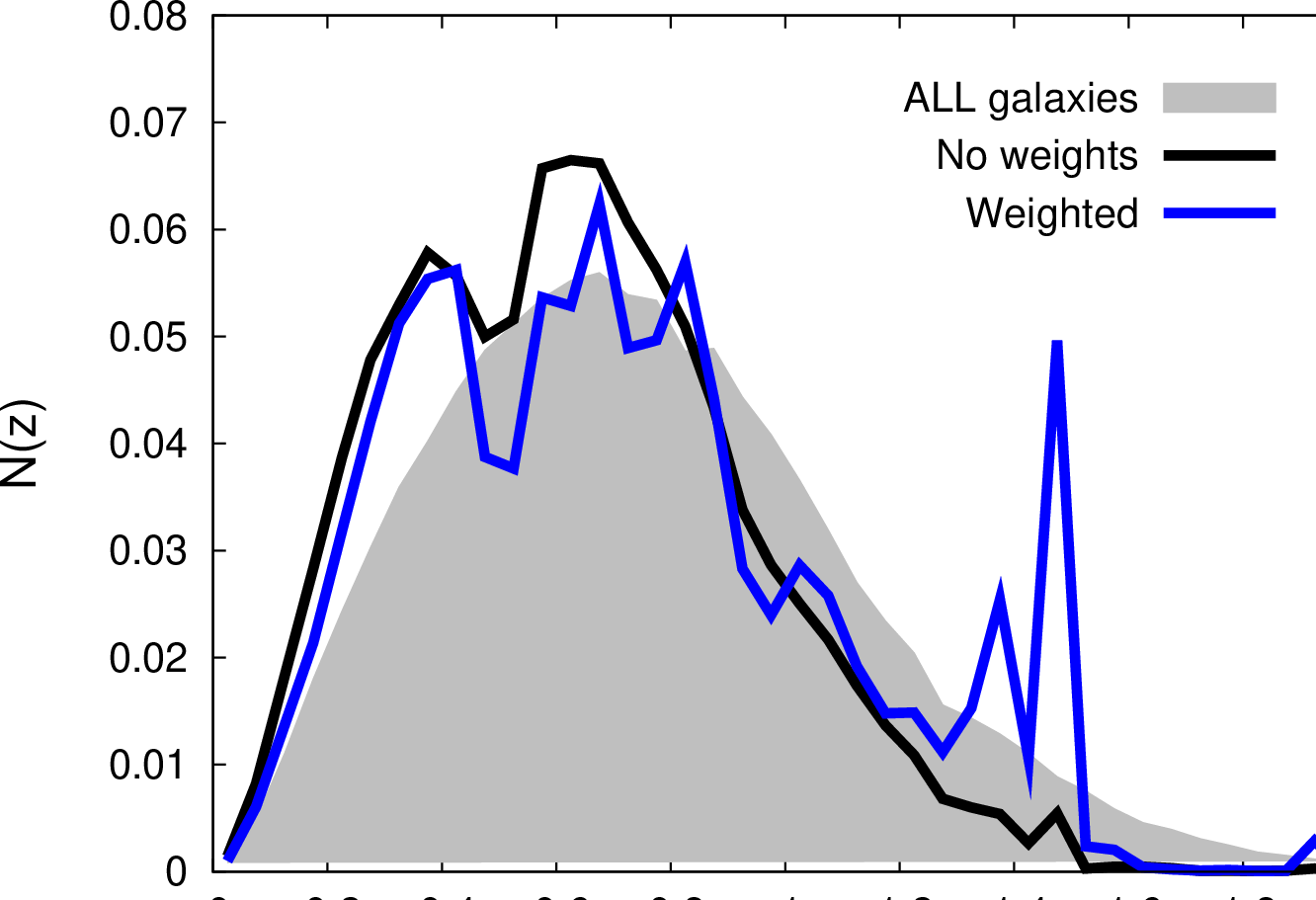}\vspace{0.5cm}
\caption{(Top plot) The true redshift distribution of the full photometric sample (shaded gray),
of the spectroscopic sample with $Q \geq 3$ with no weights (black line), and with weights (blue - dark gray line). 
(Bottom plot) Same as above, but showing weighted and unweighted distributions of 
spectroscopic redshifts.
One can see that, because wrong redshifts occupy regions of low completeness 
in observable space, the weights boost their impact enormously. 
} \label{fig:nzwei}
\end{figure}

As the top plot of Fig. \ref{fig:nzwei} shows, the weights improve the estimate 
of the overall redshift distribution when true redshifts are used.
One can see that the weights roughly fix the broadest discrepancies,
but cannot correct sharper features. 
For example, the dip in the training sample from around $0.4<z<0.6$ gets
rescaled, but its rough shape persists. 
What this suggests is that objects in this redshift range occupied the
same region of observable space, and the weighting affected them all
similarly. 

The bottom plot shows the results when the spectroscopic redshifts are used. 
We see that even a speck of wrong redshifts ($2.4\%$ in this case) can have 
dramatic impact depending on where they are located (cf. bottom plot).  
Comparing, the bottom plot of Fig. \ref{fig:nzwei}
with the middle plot of Fig. \ref{fig:zzp}, we see that the spikes in the
Weighted estimated of the redshift distribution at $z \sim$ 1.5, 1.4, 0.8, 0.7
and 0.4 all correspond to the regions of concentration of wrong redshifts seen
in Fig. \ref{fig:zzp}.  However, whereas the spikes below $z=1$ are not
particularly prominent, the spikes around $z=1.4$ and 1.5 are enormous.  There
are a couple factors contributing to the problem.  As can be inferred from the
the left plot in Fig. \ref{fig:ssr}, the completeness drops precipitously
above $z>1.4$.  Hence, the few spectroscopic redshifts above $z>1.4$ typically
receive large weights to compensate for the incompleteness.  In addition, as
shown in the middle plot of Fig. \ref{fig:zzp}, the fraction of correct
redshifts for galaxies with $z_{\rm true}>1.4$ is very small, and many of
these are incorrectly assigned a spectroscopic redshift of $\zspec=1.4$ or
1.5.  The large weights magnify the impact of the wrong redshifts, resulting
in the large spikes, and in large bias in the cosmological parameters, as we
show in the next section.

\subsubsection{Weak lensing constraints and biases with weights}

Table \ref{tab:wstatswei} shows the $1\sigma$ constraints and biases on $w$ when
one uses the weights technique to match the spectroscopic selection to the
photometric sample.  As in Sec. \ref{sec:wlres}, we separate the analysis into
two parts.  First, in the $\ztrue$ column, we show only the effect
of matching the selection between the spectroscopic and photometric samples.
Afterwards, in the $\zspec$ column, we use the actual spectroscopic
redshifts to show the impact of wrong redshifts.

As shown in the Table the weights perform reasonably for all cases
when one considers only the true redshifts.  
The biases are typically smaller than the statistical errors on $w$, 
and the statistical constraints are better than for the culling
approach of Sec. \ref{sec:wlres} since almost all of the photometric sample
was usable for analysis.  
It is interesting to note that more rigorous cuts (${\rm Q} > 6$ and 5) 
yielded the smallest biases even though the completeness of the 
spectroscopic sample was smaller than for the ${\rm Q}<4$ case.  
It is instructive to re-examine the top plot of Fig. \ref{fig:nzwei}.
Comparing the blue solid line with the shaded gray region, we see that
the weights reconstruction of the overall redshift distribution is
quite jagged, yet it yields small biases in w. 
The reason for this is that the w is sensitive to a very broad
redshift range, and an estimator can over- or under-estimate the
overall redshift distribution in small redshift ranges as long as
there are no net biases.
This point is explored in much more detailed in Sec. 5.3.1 of \citet{cun12}. 
Unfortunately, the $\zspec$ column in Table \ref{tab:wstatswei} shows that the 
presence of wrong redshifts severely compromises the weights approach.

Because the wrong redshifts are tightly associated with the regions of 
high incompleteness, particularly at high redshift, and because the variations 
in completeness are so sharp, the wrong redshifts received very large weights 
resulting in large cosmological biases.  
A major part of the problem is the sharp {\it change} in completeness with 
redshift shown on the left plot of Fig. \ref{fig:ssr}.
We find that the results for the weights do not improve for the 48600 secs 
cases because the steep variations in the completeness with redshift become 
even larger for that case since the increased exposure time did not yield 
significant increase in completeness above $z$ of 1.4.

In summary, we find that the weights approach needs to be considered with 
care in the presence of wrong redshifts, and that the more conservative 
approach of culling using the neural network is the safest. 
In practice, the weights are often needed to account for other types
of incompleteness \citep[see e.g.][]{cun09}, so both approaches should
be used in tandem.

\begin{table}
\begin{center}
\begin{tabular}{cccccc}
\hline\hline \multicolumn{6}{c}{\rule[-2mm]{0mm}{6mm} Constraints on $w$ (template-fitting photo-zs and weights) }\\
\hline\hline \multicolumn{1}{c}{\rule[-2mm]{0mm}{6mm} {\textbf{ 16200 secs}} }   &\multicolumn{3}{c}{\rule[-2mm]{0mm}{6mm} } &
\multicolumn{2}{c}{\rule[-2mm]{0mm}{6mm} bias$(w)$}\\
\hline   Selection & {\rule[-2mm]{0mm}{6mm} G. Frac.}& {\rule[-2mm]{0mm}{6mm} ${\rm SSR}_{\rm T}$ ($\%$)}  & {\rule[-2mm]{0mm}{6mm} $\sigma(w)$} & \rule[-2mm]{0mm}{6mm} $\ztrue$ & \rule[-2mm]{0mm}{6mm} $\zspec$ \\\hline
\rule[-2mm]{0mm}{6mm}$Q \geq 3$&0.73 & 93.2 &0.06& 0.070 &- 0.7     \\
\rule[-2mm]{0mm}{6mm}$Q \geq 4$&0.60 & 98.6 &0.06& 0.034 &- 0.5     \\
\rule[-2mm]{0mm}{6mm}$Q \geq 5$&0.53 & 99.6 &0.06&-0.036 &- 0.3     \\
\hline\hline
\end{tabular}
\caption{ Statistical and systematical errors in $w$ when the weights technique 
for selection matching is used.
Results are shown assuming the spectroscopic sample was selected with different cuts
of the cross-correlation strength parameter Q, described in Sec. \ref{sec:sel}. 
  The bias results
  shown used the template-fitting photo-zs.  The Gal.\ Frac.\ column indicates
  the fraction of galaxies from the {\it spectroscopic sample} that passed the
  selection cut, and the ${\rm SSR}_{\rm T}$ indicates the fraction of correct redshifts 
  (i.e, fraction for which $|\zspec-\ztrue|<0.01$) in the sample.  
  Essentially all of the photometric sample was used in the analysis,
  hence the statistical constraints are the same for all samples.}
\label{tab:wstatswei}
\end{center}
\end{table}

\section{Discussion: Robustness of assumptions and results}\label{sec:robust}

We now discuss the dependence of our results on the key assumptions 
and numerical tools used in this work.

\begin{itemize}

\item {\it N-body/photometric simulations}: The success rate statistics are
  affected by luminosity function and distribution of galaxy types in the
  simulation.  However, the main conclusions of our paper, concerning
  selection matching and impact of wrong redshifts, should not be affected.  We
  tested the selection matching for a variety of situations (several of which
  we do not show), including varying atmospheric noise models and spectrograph
  resolution.  For all cases, the matching worked well, incurring no
  additional bias.  In addition, Soumagnac et al. (in preparation) obtain
  similar results using a very different set of spectro/photometric
  simulations described in \cite{jou09}.

The distribution of wrong redshifts in $(\ztrue,\zspec)$ space could also change
for a different simulation, but the preferred loci where the failures concentrate
should not vary appreciably, since they are based on confusion between galaxy or
atmospheric spectral lines that do not depend on  any  details of the simulation.
Furthermore, the fact that a small fraction of spectroscopic 
failures can cause severe biases is not likely to change.

\item {\it Sky noise model:} Our model for sky subtraction is idealized as it
  assumes a perfect shot-noise model.  Sky-subtraction is often not as
  efficient, and observing conditions vary from the median.  In addition,
  there are issues such as CCD fringing (cf. Sec. \ref{sec:addsys}) which are
  difficult to model.  Other effects we did not model include contamination
  from nearby stars or bright galaxies, and cosmic rays.
  These other effects, however, are only expected to affect the overall 
  completeness, without galaxy type or redshift dependence.

\item {\it Simulated spectra:} As discussed in Appendix \ref{sec:pipe}, the
  simulated spectra we use are based on the 5 eigenspectra of {\tt kcorrect},
  which are derived based on about 1600 SDSS main sample galaxies, 400
  luminous red galaxies and a photometric sample of several thousands of
  galaxies imaged in the UV, optical and IR.  Is this enough?  \cite{yip04}
  showed that a set of 3 eigentemplates were sufficient to describe about
  $98\%$ of the variance in the 170,000 galaxies in the \cite{str02} SDSS
  sample.  Additional templates improved coverage very slowly, with a set of
  500 eigentemplates needed to account for $99\%$ of the sample variance
  (cf. Table 1 in that work).  \cite{yip04} show that the missing variance was
  due mainly to extreme line-emission galaxies.  We roughly confirm this trend
  for our simulated spectra by looking at the distribution of equivalent widths
  of the [OII] emission line for our simulated galaxies.  We find that our
  equivalent widths reach at most 30 ${\rm \AA}$.  For comparison, \cite{coo06} find,
  for the DEEP2 sample, a distribution of [OII] equivalent widths reaching as 
  much as 100 ${\rm \AA}$.

In addition, \cite{yip04} showed that one needs a random subsample of about
10,000 galaxies to obtain convergence for the first 10 eigentemplates.  These
results suggest the {\tt kcorrect} basis should be sufficient to characterize
all but a few percent of the low-redshift galaxies\footnote{The \cite{yip04}
  analysis was based on principal component analysis, whereas \cite{bla07}
  used non-negative matrix factorization to determine their respective
  eigenbasis.  Thus, comparison between \cite{bla07} and \cite{yip04} are only
  meant as ballpark estimates.}.  However, a few percent of ``oddball''
galaxies could potentially cause problems for cosmological analysis if they
cannot be disentangled from the rest of the sample using colors and if their
redshift distribution differs significantly from the rest of the sample with
similar colors.  The problem is expected to become worse at high-redshift.  To
properly quantify the impact of the outliers, observing campaigns targeted at
the spectroscopic failures of existing spectroscopic surveys are crucial.

In some sense, our choice of template library used for deriving
spectroscopic redshifts is pessimistic for the high-redshift galaxies: as discussed in 
Appendix \ref{sec:pipe}, the {\tt kcorrect} templates are based on GALEX colors 
for the bluer frequencies.
Hence, parts of the spectra of high-z objects were simulated using purely
photometric data, resulting in excessively featureless spectra in the UV
frequencies, which implied lower-than-expected completeness for $z>1.4$.

\item {\it Spectroscopic redshift pipeline} The {\tt rvsao.xcsao} code uses
cross-correlation techniques in Fourier space to derive redshifts from spectra.
The disadvantage of this approach relative to a standard $\chi^2$ method is that 
one does not include any information about the noise.
One can disregard certain regions of the spectrum in the analysis, thereby removing
at least the most prominent atmospheric lines. 
We found that the removal of some lines did not increase the completeness of the sample
noticeably, and changed the distribution of the wrong redshifts.
We leave more extensive tests on the optimal techniques for spectroscopic redshift 
estimation for a future work.

\item {\it Culling approach to selection matching} In a real
  spectroscopic survey, the success of the neural network culling approach to
  selection matching is likely to require a more careful redshift confidence
  classification. In our simulation, the atmospheric conditions were fixed for
  the full sample, and the noise subtraction was perfect. In a real survey,
  the variability of observing conditions and other problems such as slit
  placement and cosmic ray contamination will complicate the mapping between
  magnitudes in the photometric surveys and the redshift confidence in the
  spectroscopic survey. This difficulty can likely be overcome if some measure
  of the observing conditions is used as input when training the neural
  network. We leave these investigations for a future work.




\end{itemize}

\section{Implications for Survey Design}\label{sec:design}

Given the findings of this paper and \cite{cun12}, what should survey planners
do to optimize their spectroscopic surveys?

The first step is obvious: one needs to optimize the allocation of time
observing different kinds of galaxies. Specifically, one can use color
  information to preselect galaxies that will require longer exposure times to
  obtain accurate redshifts. For example, in Sec.~\ref{sec:wlres}, we saw
that tripling the exposure time improved the completeness from 0.46 to 0.66
for the $\qest>3.5$ cut.  If the $20\%$ of the sample that yielded additional
redshifts could be known in advance, one would only target this sample for
additional observation, which would only require an increase of $40\%$ in the
observing time, instead of the naive $200\%$ additional time if the full
sample was targeted for follow-up observation. 
With an optimized observing strategy, one would be able to save precious
telescope time and still achieve redshift accuracy that does not degrade
the cosmological constraints appreciably. We leave a more detailed analysis for
future work.

We showed in this paper that the tolerance for wrong redshifts is extremely
low.  It is, however, possible to get away with a higher fraction of wrong
spectroscopic redshifts by modeling their effects on the cosmological
parameters.  
Then one would need to, in analogy to the photo-z case, fully characterize 
the spectroscopic error matrix $\pzst$. 
However, determining  the matrix $\pzst$ from observations is likely to be 
very challenging in  practice, as in order to control the sample variance of 
galaxies used for  the calibration, one would likely have excessively high 
requirements on the  area of the follow-up \citep{cun12}.

It is also possible that one can use spatial cross-correlations to estimate
the spectroscopic error matrix.  Since correlations between different redshift
bins should be very close to zero, any correlation has to be due to wrong
redshifts.  Several works have explored this fact for photo-z calibration
\citep{sch06,erb09,ben10,zha10}.  \cite{sch06}, for example, found
cross-correlations to work well only in the simplest Gaussian cases.  But for
spectroscopic failures, the excess correlation signal should be due to a few
big outliers, and hence might be more easily detectable.

\section{Conclusions}\label{sec:concl}

We investigated the impact of spectroscopic failures on the training and
calibration of photometric redshifts, and the consequent impact on the
forecasted dark energy parameter constraints from weak gravitational lensing.
Our tests were based on N-body/spectrophotometric simulations patterned after
the DES and expected spectroscopic follow-up observations loosely patterned
after the VVDS survey.

Spectroscopic failures consist of two types of issues: the inability to obtain
spectroscopic redshifts for certain galaxies, and incorrect redshifts.

The inability to obtain redshifts introduces incompleteness in the spectroscopic
sample --- i.e. missing redshifts in some region of parameter space (e.g.\ at
faint magnitudes) represented in the full photometric population of galaxies.
This incompleteness must be accounted for before one can use the spectroscopic
sample to calibrate photo-zs -- i.e characterize the photo-z error matrices,
e.g. the $\pzs$, of the sample.  

We studied two approaches to account for the incompleteness in the
spectroscopic sample.  In the first approach, we used an artificial neural
network to estimate the spectroscopic selection function for the photometric
sample.  This selection function was then used to cull the photometric sample
so that its statistical properties matched the spectroscopic sample.  We found
this approach works extremely well, yielding only insignificant bias in the WL
constraints using the culled sample (refer to $\ztrue$ column in Table
\ref{tab:wstats}).  However, the statistical constraints did degrade
substantially as, typically, a large fraction of the sample was culled.  In
the second approach, we accounted for the incompleteness in the spectroscopic
sample by applying weights to the galaxies with spectroscopic redshifts,
following the approach of \cite{lim08}, so that the statistical properties of
the spectroscopic and photometric samples match.  This approach was also
successful (cf. $\ztrue$ column in Table \ref{tab:wstatswei}) --- as expected,
because most of the photometric sample could be used --- yielding tolerable
cosmological biases while obtaining the maximum statistical constraints.
Overall, we found that the effects of spectroscopic incompleteness are well
under control.

Unfortunately, on the other hand, we found that wrong redshifts can
significantly degrade cosmological constraints and $>99\%$ of correct
spectroscopic redshifts seems to be needed (cf.\ $\ssrt$ and $\zspec$ columns
in Tables \ref{tab:wstats} and \ref{tab:wstatswei}).  We found the results to
be independent of the photo-z estimators used, but somewhat dependent on the
settings of the spectroscopic pipeline.  In particular, we found that attempts
to increase the completeness of the spectroscopic sample during the spectral
analysis can result in more catastrophic spectroscopic redshift failures,
which will increase cosmological biases.

We tested a couple of approaches to identify wrong spectroscopic redshifts,
finding that the NNE error estimator \citep{oya08b} is able to reduce the bias
in the measured dark energy equation of state by half while removing only
$10\%$ of the photometric sample.  Slightly less improvement in the $w$ bias
was obtained using the template-fitting error estimator.

In summary, we find that wrong redshifts are by far the main issue affecting
calibration of photo-z error distributions with spectroscopic samples.  Future
follow-up spectroscopic observations of the planned and ongoing wide-area
photometric surveys must focus primarily on the accuracy of the spectroscopic
redshifts even if that implies sacrificing the spectroscopic completeness.

\section*{Acknowledgments}

CEC would like to thank Joerg Dietrich, Stephanie Jouvel, Anja von der Linden, 
Jeff Newman and Peter Norberg for discussions about spectroscopic surveys.
We also thank G. Bernstein, J. Cohn, and S.\ Lilly for detailed comments on 
the draft.
This paper has gone through internal review by the DES collaboration.
CEC is supported by a Kavli Fellowship at Stanford University. DH is supported
by the DOE OJI grant under contract DE-FG02-95ER40899.  DH is additionally
supported by NSF under contract AST-0807564, and NASA under contract
NNX09AC89G.  RHW received support from the U.S. Department of Energy under
contract number DE-AC02-76SF00515.  MTB was supported by Stanford University
and the Swiss National Science Foundation under contract 2000 124835/1.  This
research was supported in part by the National Science Foundation under Grant
No. PHY05-51164, Grant No. 1066293 and the hospitality of the Aspen Center for
Physics.  Fermilab is operated by Fermi Research Alliance, LLC under Contract
No.  DE-AC02-07CH11359 with the United States Department of Energy.

\appendix
\section{The simulations} \label{app:sims}

In this section, we describe the construction of the simulations used in our analysis.

\subsection{N-body/photometric simulations}\label{sec:nbody}

The simulated galaxy catalog used for the present work was generated
using the Adding Density Determined GAlaxies to Lightcone Simulations
(ADDGALS) algorithm \citep{wechsler-addgals,busha-addgals}.  This
algorithm attaches synthetic galaxies to dark matter particles in a
lightcone output from a dark matter N-body simulation.  The model is
designed to match the luminosities, colors, and clustering properties
of galaxies.

The simulations used here start with a dark matter lightcone which
spans the redshift range from $0 < z < 2$, over one octant of sky
(5156 sq. degrees).   The lightcone is constructed from three distinct
N-body simulations, which range in resolution from a few $10^{10}$ to
a few $10^{11} \Msun$ particles and box sizes ranging from 1 to 4 Gpc/$h$.   
The simulations were run with the LGadget code and modeled a flat $\Lambda
CDM$ cosmology using parameters consistent with WMAP7 results.

The ADDGALS algorithm used to create the galaxy distribution consists
of two steps: galaxies based on an input luminosity function are first
assigned to particles in the simulated lightcone, after which
multi-band photometry is added to each galaxy using a training set of
observed galaxies.  For the first step, we begin by defining the
relation $P(\delta_{dm}|M_r, z)$ --- the probability that a galaxy
with magnitude $M_r$ a redshift $z$ resides in a region with local
density $\delta_{dm}$, defined as the radius of a sphere containing
$1.8 \times 10^{13} \hinv\Msun$ of dark matter.  This relation can be
tuned to reproduce the luminosity-dependent galaxy 2-point function by
using a much higher resolution simulation combined with the technique
known as subhalo abundance matching.  This is an algorithm for
populating very high resolution dark matter simulations with galaxies
based on halo and subhalo properties that accurately reproduces
properties of the observed galaxy clustering \citep{conroy06,
  wetzel10, behroozi10, busha11}.  The relationship
$P(\delta_{dm}|M_r,z)$ can be measured directly from the resulting
catalog.  Once this probability relation has been defined, galaxies
are added to the simulation by integrating a (redshift dependent)
$r$-band luminosity function to generate a list of galaxies with
magnitudes and redshifts, selecting a $\delta_{dm}$ for each galaxy by
drawing from the $P(\delta_{dm}|M_r, z)$ distribution, and attaching
it to a simulated dark matter particle with the appropriate
$\delta_{dm}$ and redshift.  The advantage of ADDGALS over other
commonly used approaches based on the dark matter halos is the ability
to produce significantly deeper catalogs using simulations of only
modest size.  When applied to the present simulation, we populate
galaxies as dim as $M_r \approx -14$, compared with the $M_r \approx
-21$ completeness limit for a standard halo occupation (HOD) approach.

While the above algorithm accurately reproduces the distribution of
satellite galaxies, central objects require explicit information about
the mass of their host halos.  Thus, for halos with more than 100
particles, we assign central galaxies using the explicit
mass-luminosity relation determined from our calibration catalog.  We
also measure $\delta_{dm}$ for each halo, which is used to draw a
galaxy from the integrated luminosity function with the appropriate
magnitude and density to place at the center.

For the galaxy assignment algorithm, we choose a luminosity function
that is similar to the SDSS luminosity function as measured in
\cite{bla03}, but evolves in such a way as to reproduce the higher
redshift observations (e.g., SDSS-Stripe 82, AGES, GAMA, NDWFS and
DEEP2).  In particular, $\phi_*$ and $M_*$ are varied as a function of 
redshift in accordance with the recent results from GAMA \citep{lov12}.

Once the galaxy positions have been assigned, photometric properties
are added.  We begin with a training set of spectroscopic galaxies and
the simulated set of galaxies with $r$-band magnitudes generated
earlier.  For each galaxy in both the training set and simulation we
measure $\Delta_5$, the distance to the 5th nearest galaxy on the sky
in a redshift bin.  Each simulated galaxy is then assigned an SED
based on drawing a random training-set galaxy with the appropriate
magnitude and local density, k-correcting to the appropriate redshift,
and projecting onto the desired filters.  When doing the color assignment,
the likelihood of assigning a red or a blue galaxy is smoothly varied as a 
function of redshift in order simultaneously reproduce the observed red
fraction at low and high redshifts as observed in SDSS and DEEP2.  

Differences between the training set and simulated galaxy sample
complicate the process of color-assignment.  In order to compile a
sufficiently large training set, we use a magnitude-limited sample of
SDSS spectroscopic galaxies brighter than $m_r = 17.77$ with $z <
0.2$.  The simulated sample, on the other hand, is a volume-limited
sample, spanning a broader redshift range.  When measuring $\Delta_5$
we restrict ourselves to neighbors brighter than $M_r = -19.7$ in the
simulation sample, while using all objects in the observational
catalog.  To mitigate differences in luminosity and redshift,
each galaxy is rank ordered according to its density in its redshift bin,
and require that objects be in the same percentile bin in each sample
rather than having the same the absolute value of $\Delta_5$.  This is
similar to the method used in \cite{cooper08}.

The final step for producing a realistic simulated catalog is the
application of photometric errors.  While the photometric errors
generated here are particular to DES, the algorithm can be generalized
for any survey.  For each galaxy, we add a noise term to the intrinsic
galaxy flux, where the noise is drawn from a Gaussian of width
\begin{equation}
{\rm noise} = \sqrt{t_e n_p n_s + f_{g,i} t_e}
\end{equation}
where $t_e$ is the exposure time, $n_p$ the number of pixels covered by a
galaxy, $n_s$ the flux of the sky in a single detector pixel, and $f_{g,i}$ is
the intrinsic flux of the galaxy.  Here, galaxies are assumed to have the same
angular size, hence $n_p$ is identical for all objects.  Application of the
above relation to objects from the SDSS catalog shows that it is able to
faithfully reproduce the reported errors of the survey.

\subsection{Creating simulated spectra}\label{sec:pipe}

We use the {\tt kcorrect v4\_1} code \citep{bla03} to derive simulated
spectra.  The {\tt kcorrect} code includes a set of 5 eigenspectra derived
using a non-negative matrix factorization (NMF) technique \citep{bla07}.  To
derive the eigenspectra, the authors start out with a basis of 450 star
formation history templates from \cite{bru03} as well as 35 templates from
\cite{kew01}.  The method uses this basis to derive the nonnegative linear
combination of templates that best described the observations.  In this case,
the observations consist of a sample of several thousand photometrically
and/or spectroscopically observed galaxies, from the far UV to the near IR
\citep{bla07}.  The spectroscopic part of the training data consisted of 400
SDSS luminous red galaxies (LRGs) with $0.15<z<0.5$ \citep{eis01} and 1600
SDSS main sample galaxies with $0.0001<z<0.4$ \citep{str02}, with both sets of
data observed in the range $3800{\rm \AA}<\lambda<9000{\rm \AA}$.

We use the {\tt kcorrect} subroutine to convert the true redshift and
error-free magnitudes of a simulated galaxy from our photometric simulation
into a best-fitting spectral energy distribution (SED).  The SED is
characterized by the coefficients of the 5 eigentemplates, and are output as
the variable {\tt coeffs}.  The {\tt coeffs} are then passed into the
subroutine {\tt k\_reconstruct\_spec}, which produces a simulated spectrum
with a resolution, in units of velocity dispersion, of 300 km/s.

We pattern our mock survey loosely on the VIMOS-VLT Deep Survey \citep[VVDS;][]{lef05}. 
The characteristics of the instrument that we assume are: collecting area of $16 \pi$ ${\rm m^2}$, 
aperture of $5\times 0.5 \ {\rm arcsecs^2}$.
For simplicity, we assume a constant resolution and a dispersion of 
$\Delta \lambda=7.14/{\rm pixel}$ over the entire spectrograph range of $5500-9500 {\rm \AA}$.  
Comparing the spectrograph window of $5500-9500 {\rm \AA}$ \ to the
spectroscopic coverage of the training set used to create the simulated
spectra, we see that for objects below redshift of 0.05, there is no
spectroscopic representation of the training set galaxies in the range
$9000-9500 {\rm \AA}$.  More problematic is the fact that the spectroscopic
training set has wavelength coverage starting at $3800 {\rm \AA}$, and only
goes to $z=0.4$.  As a result, for galaxies at about $z>1.0$, the blue side of
the simulated spectra are based solely on photometric data.  Considering that
most of the SDSS main sample is below redshift of 0.2, the simulated spectra
should begin to lose resolution in the blue-end for $z>0.73$.  These
limitations in the simulated spectra result in higher-than-expected
incompleteness above $z=1.4$, but do not affect the overall conclusions.

\begin{figure}
\centering
  \includegraphics[scale=0.35,angle=-90]{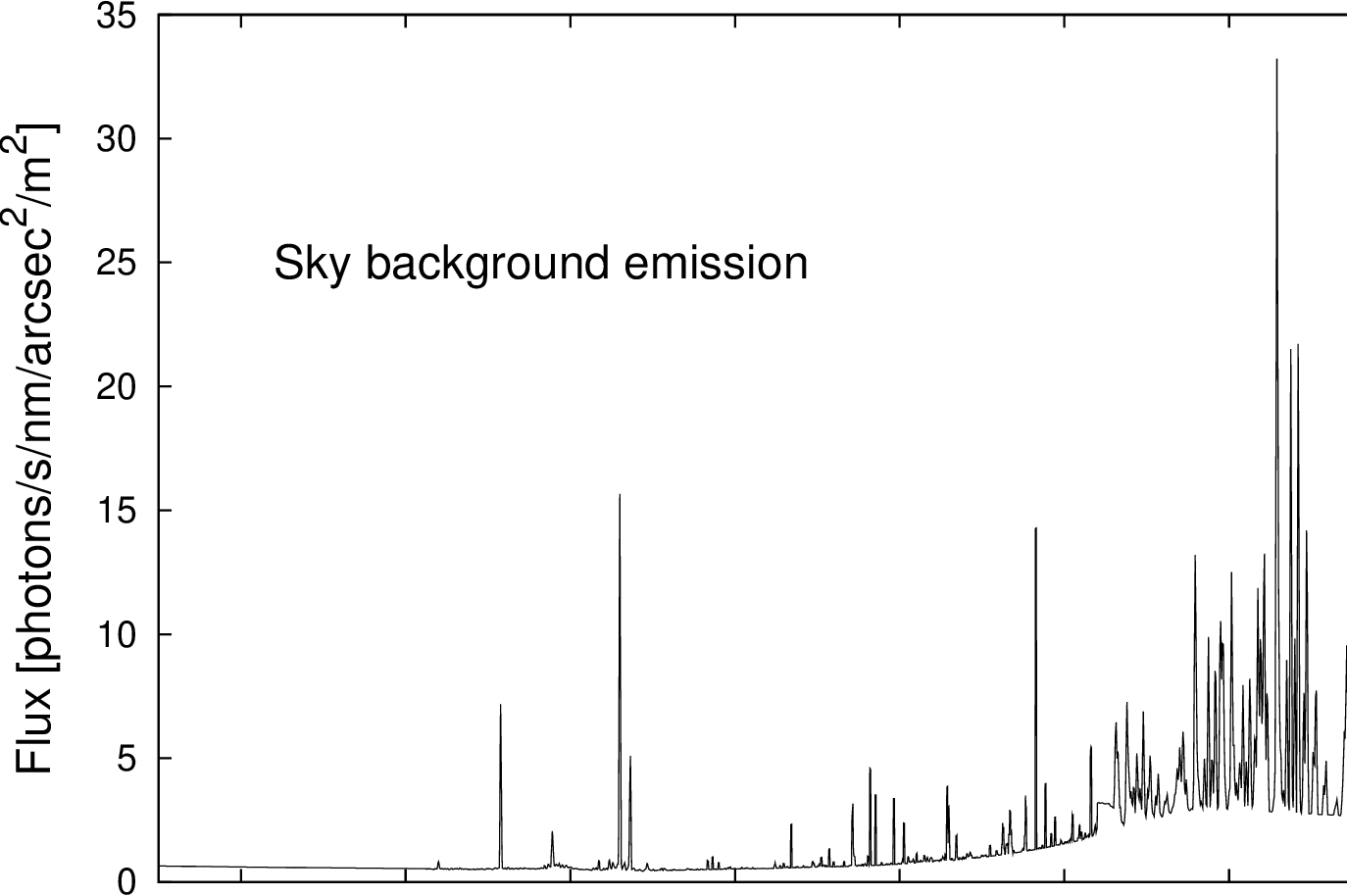}\vspace{0.8cm}
  \includegraphics[scale=0.35,angle=-90]{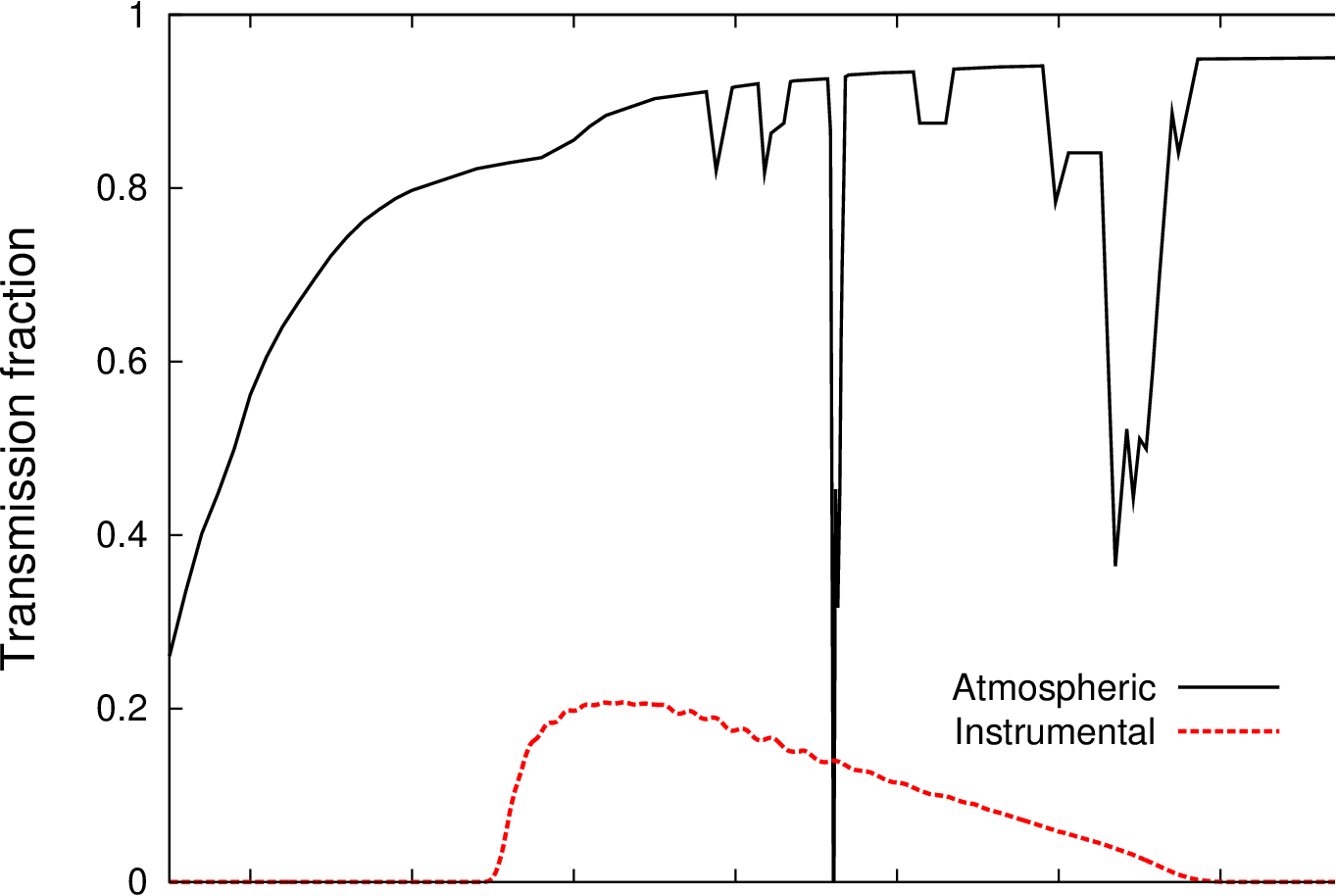}\vspace{0.5cm}
\caption{Top panel: Atmospheric emission in units of photons/s/nm/${\rm m^2}$/${\rm arcsec^2}$. 
Bottom panel: Atmospheric and instrumental transmission fractions, i.e fraction of photons that 
reach the focal plane, used in our simulation. 
The total transmission function is given by the product of the two transmissions.} \label{fig:pipe}
\end{figure}

We use a Palomar sky extinction model (courtesy of B. Oke and J. Gunn) with 1.3 airmasses and 
altitude of 2635 meters to calculate the atmospheric transmission
fraction (the solid black line in the bottom panel of Fig. \ref{fig:pipe}). 
The instrument transmission is based on the VIMOS instrument transmission 
function\footnote{\url{http://www.eso.org/observing/etc/bin/gen/form?INS.NAME=VIMOS+INS.MODE=SPECTRO}} 
and is shown as the dashed red line in the bottom panel of Fig. \ref{fig:pipe}. 
The total transmission is the product of the atmospheric and instrumental transmissions.
We assume 16200 secs exposures for the fiducial observation strategy and
also investigate a scenario with 48600 secs exposures.

We add atmospheric emission based on the sky spectrum\footnote{Sky spectrum obtained from 
{\url{ http://www.gemini.edu/sciops/ObsProcess/obsConstraints/atm-models/skybg\_50\_10.dat}}}
shown at the top panel of Fig. \ref{fig:pipe}. 
The total noise is given by the rms sum of the atmospheric noise, shot-noise from 
the galaxy spectrum itself and readout noise per pixel, which we take to be a 
constant 5 photons.
In reality, we only simulate the sky-subtracted spectrum, as follows.
First, we convert the different spectra into photon counts for each pixel.
We then assume the atmospheric and galaxy noise follow a Poisson distribution, 
so that the uncertainty in the produced noise is the square-root of the number 
of photons emitted.
The readout noise is taken to be Gaussian. 
We calculate the total noise, $N$ as 
\begin{equation}
N=\sqrt{n_{\rm atm} + n_{\rm gal} + n^2_{\rm read}}
\end{equation}
\noindent where $n_{\rm atm}$, $n_{\rm gal}$, and $n_{\rm read}$ are the number 
of photons from the atmosphere, the galaxy and the readout noise, respectively.
The expected signal is simply the total number of photons from the galaxy.
The expectation value of the error in the flux, $\delta F$ is then given by
\begin{equation}
\delta F=F\frac{N}{S}
\end{equation}
To obtain the sky-subtracted galaxy spectrum we, at each pixel, sample from a Gaussian 
distribution with mean given by the flux and width given by the error in the flux $\delta F$.

\section{Artificial Neural Networks}\label{sec:neunet}

We use an Artificial Neural Network (ANN) method to both estimate
the spectroscopic redshift quality and photometric redshifts, using
an implementation based on \citep{col04, oya08a}
Despite the fancy name, an ANN is simply a function 
which relates redshifts (or any quantity we wish to estimate) to photometric 
observables. 
The training set is used to determine the best-fit value 
for the free parameters of the ANN.
The best-fit parameters are found by minimizing the overall scatter 
 of the photo-zs determined for the training set galaxies.
The ANN configurations are not unique in the sense that different sets of 
parameters can result in the same overall scatter. 
The best-fit parameters found after minimizing the scatter depend on where in 
parameter space the optimization run begins.
Hereafter we refer to an ANN function using a given set of best-fit 
parameters as a neural network solution.

The technical details are as follows.
We use a particular type of ANN called a Feed Forward Multilayer
Perceptron (FFMP), which 
consists of several nodes arranged in layers through which 
signals propagate sequentially. 
The first layer, called the input layer, receives the input photometric 
observables (magnitudes, colors, etc.). 
The next layers, denoted hidden layers, propagate signals until 
the output layer, whose outputs are the desired quantities, in this
case the photo-z estimate or the redshift quality Q estimate. 
Following the notation of \cite{col04}, we denote a network with 
$k$ layers and $N_i$ nodes in the $i^{th}$ layer as $N_1:N_2: ... :N_k$. 

A given node can be specified by the layer it belongs to and the 
position it occupies in the layer. Consider a node in layer $i$ and 
position $\alpha$  with $\alpha=1,2,...,N_i$. 
This node, denoted $P_{i\alpha}$, receives
a total input $I_{i\alpha}$ and fires an output $O_{i\alpha}$ given by
\begin{eqnarray}
O_{i\alpha}=F(I_{i\alpha}) \,,
\end{eqnarray}  
where $F(x)$ is the activation function. 
The photometric observables are the inputs $I_{1\alpha}$ to the 
first layer nodes, which produce outputs $O_{1\alpha}$. 
The outputs $O_{i\alpha}$ in layer $i$ are
propagated to nodes in the next layer $(i+1)$, denoted $P_{(i+1)\beta}$,
with $\beta=1,2,..N_{i+1}$. 
The total input $I_{(i+1)\beta}$ is a weighted sum of the outputs 
$O_{i\alpha}$
\begin{eqnarray}
I_{(i+1)\beta} = \sum_{\alpha=1}^{N_i} w_{i\alpha\beta} O_{i\alpha},
\end{eqnarray}
where $w_{i\alpha\beta}$ is the weight that connects nodes 
$P_{i\alpha}$ and $P_{(i+1)\beta}$.
Iterating the process in layer $i+1$, signals propagate from hidden layer 
to hidden layer until the output layer.
In our implementation, we use a network configuration $N_m:10:10:10:1$, 
which receives $N_m$ magnitudes and outputs a photo-z or a spectroscopic redshift quality. 
We use hyperbolic tangent activation  functions in the hidden layers and a 
linear activation function for the output layer.

\section{Biases in weak lensing measurements of dark energy}\label{sec:wl_app}

To assess the impact of the spectroscopic failures on the cosmological
parameters, we closely follow the formalism used in our previous work on the
impact of sample variance to photo-z calibration \citep{cun12}. 
We consider a weak lensing survey,
and for simplicity only study the shear-shear correlations.  The observable
quantity we consider is the convergence power spectrum
\begin{equation}
C^{\kappa}_{ij}(\ell)=P_{ij}^{\kappa}(\ell) + 
\delta_{ij} {\langle \gamma_{\rm int}^2\rangle \over \bar{n}_i},
\label{eq:C_obs}
\end{equation}
where $\langle\gamma_{\rm int}^2\rangle^{1/2}$ is the rms intrinsic
ellipticity in each component, $\bar{n}_i$ is the average number of galaxies
in the $i$th redshift bin per steradian, and $\ell$ is the multipole that
corresponds to structures subtending the angle $\theta = 180\degree/\ell$.
For simplicity, we drop the superscripts $\kappa$ below.  We take
$\langle\gamma_{\rm int}^2\rangle^{1/2}=0.26$.

We  follow the formalism of \citet{BH10} (hereafter BH10), where the
photometric redshift errors are algebraically propagated into the biases in
the shear power spectra. These biases in the shear spectra can then be
straightforwardly propagated into the biases in the cosmological parameters.
We now review briefly this approach.

Let us assume a survey with the (true) distribution of source galaxies in
redshift $n_t(z)$, divided into $B$ bins in redshift. Let us define the
following terms
\begin{itemize}
\item {\em Leakage} $P(z_p|z_t)$ (or $l_{tp}$ in BH10 terminology): fraction
  of objects from a given true redshift bin that are placed into an incorrect
  (non-corresponding) photometric bin.

\item {\em Contamination} $P(z_t|z_p)$ (or $c_{tp}$ in BH10 terminology): fraction of galaxies
  in a given photometric bin that come from a non-corresponding true-redshift
  bin.
\end{itemize}

When specified for each tomographic bin, these two quantities contain the same
information.  Note in particular that the two quantities satisfy the
integrability conditions
\begin{eqnarray}
\int P(z_p|z_t)dz_p &\equiv& \sum_p l_{tp} = 1\\[0.2cm]
\int P(z_t|z_p)dz_t &\equiv& \sum_t c_{tp} = 1.
\end{eqnarray}

A fraction $l_{tp}$ of galaxies in some
true-redshift bin $n_t$ ``leak'' into some photo-z bin $n_p$, so that
$l_{tp}$ is the fractional perturbation in the true-redshift bin, while the
contamination $c_{tp}$ is the fractional perturbation in the photometric bin.
The two quantities can be related via
\begin{equation}
c_{tp} = \frac{N_t}{N_p}\, l_{tp} 
\label{eq:contamination}
\end{equation}
where $N_t$ and $N_p$ are the absolute galaxy numbers in the true
and photometric redshift bins, respectively. Then,
\begin{eqnarray}
n_t &\rightarrow & n_t\\[0.1cm]
n_p &\rightarrow & (1-c_{tp})\,n_p + c_{tp}\,n_t
\label{eq:n_bias}
\end{eqnarray}
and the photometric bin normalized number density is affected (i.e. biased) by
photo-z catastrophic errors.  The effect on the cross power spectra is then
\begin{eqnarray}
C_{pp} &\rightarrow &(1-c_{tp})^2C_{pp} +2c_{tp} (1-c_{tp})C_{tp} + c_{tp}^2 C_{tt}
           \nonumber\\[0.1cm]
C_{m p} &\rightarrow &(1-c_{tp})C_{mp} +c_{tp}\, C_{mt} 
           \qquad (m< p)
	   \label{eq:C_bias}\\[0.1cm]
C_{p n} &\rightarrow &(1-c_{tp})C_{pn} +c_{tp}\, C_{tn} 
           \hspace{0.9cm} (p < n)\nonumber \\[0.1cm]
C_{mn} &\rightarrow &C_{mn}\qquad\qquad\qquad\qquad\qquad  ({\rm otherwise})
	   \nonumber
\end{eqnarray}
(since the cross power spectra are symmetrical with respect to the interchange
of indices, we only consider the biases in power spectra $C_{ij}$ with $i\leq j$).  
Note that these equations are exact for a fixed contamination coefficient $c_{tp}$. 

The bias in the observable power spectra is the rhs-lhs difference in the
above equations\footnote{We have checked that the quadratic terms in $c_{tp}$
  are unimportant, but we include them in any case.}.  The cumulative result
due to all contaminations in the survey (or, $P(z_t|z_p)$ values for each
$z_t$ and $z_p$ binned value) can be obtained by the appropriate sum
\begin{eqnarray}
\delta C_{pp} &=&\sum_t (-2c_{tp} + c_{tp}^2)C_{pp} +2c_{tp} (1-c_{tp})C_{tp} + c_{tp}^2 C_{tt}
           \nonumber\\[0.1cm]
\delta C_{mp} & =&\sum_t \left (-c_{tp}C_{mp} +c_{tp}\, C_{mt}\right )
  \label{eq:delta_Cmp}\\[0.1cm]
\delta C_{pn} & =&\sum_t \left (-c_{tp}C_{pn} +c_{tp}\, C_{tn}\right )  \nonumber
\end{eqnarray}
for each pair of indices $(m, p)$, where the second and third line assume
$m<p$ and $p<n$, respectively.

The
bias in cosmological parameters is given by using the standard linearized
formula \citep{Knox_Scocc_Dod,Huterer_Turner},  summing over each pair of
contaminations $(t, p)$
\begin{equation}
\delta p_i \approx \sum_{j}(F^{-1})_{ij} \sum_{\alpha\beta} {\partial \bar
  C_\alpha \over \partial p_j} ({\rm Cov}^{-1})_{\alpha\beta}\,\delta C_\beta ,
\label{bias1}
\end{equation}
where $F$ is the Fisher matrix and ${\rm \bf Cov}$ is the covariance of shear
power spectra (see just below for definitions). This formula is accurate when
the biases are 'small', that is, when the biases in the cosmological
parameters are much smaller than statistical errors in them, or $\delta p_i
\ll (F^{-1})_{ii}^{1/2}$.  Here $i$ and $j$ label cosmological parameters, and
$\alpha$ and $\beta$ each denote a {\it pair} of tomographic bins,
i.e.\ $\alpha, \beta=1, 2, \ldots, B(B+1)/2$, where recall $B$ is the number
of tomographic redshift bins.  To connect to the $C_{mn}$ notation in
Eq.~(\ref{eq:C_bias}), for example, we have $\beta = mB + n$.

We calculate the Fisher matrix $F$ assuming perfect redshifts, and following
the procedure used in many other papers \citep[e.g.][]{MMG}. 
The weak lensing Fisher matrix is then given by
\begin{equation}
F^{\rm WL}_{ij} = \sum_{\ell} \,{\partial {\bf C}\over \partial p_i}\,
{\bf Cov}^{-1}\,
{\partial {\bf C}\over \partial p_j},\label{eq:latter_F}
\end{equation}
where $p_i$ are the cosmological parameters and ${\bf Cov}^{-1}$ is
the inverse of the covariance matrix between the observed power spectra whose
elements are given by
\begin{eqnarray}
{\rm Cov}\left [C_{ij}(\ell), C_{kl}(\ell')\right ] &=& 
{\delta_{\ell \ell'}\over (2\ell+1)\,f_{\rm sky}\,\Delta \ell} \times \\
&&\left [ C_{ik}(\ell) C_{jl}(\ell) + 
  C_{il}(\ell) C_{jk}(\ell)\right ]. \nonumber 
\label{eq:Cov}
\end{eqnarray}
where $\Delta\ell$ is the width of the binning in
multipole $\ell$. The
survey specifications and the cosmological parameter set is described in
Sec.~\ref{sec:wl}. Finally, we add the information expected from the Planck
survey given by the Planck Fisher matrix (W.\ Hu, private communication). The
total Fisher matrix we use is thus
\begin{equation}
F = F^{\rm WL}+F^{\rm Planck}.
\label{eq:Fisher}
\end{equation}
The fiducial constraint on the equation of state of dark energy assuming
perfect knowledge of photometric redshifts is $\sigma(w)=0.055$.

Our goal is to estimate the biases in the cosmological parameters due to
imperfect knowledge of the photometric redshifts. In particular, the relevant
photo-z error will be the difference between the inferred $P(z_s|z_p)$
distribution for the calibration (or, training) set -- using spectroscopic redshifts
as a proxy for the true redshifts -- and the $P(z_t|z_p)$ distribution for the actual survey.  
Therefore, we define 
\begin{eqnarray}
\delta C_{ij} &=& C_{ij}^{\rm train} - C_{ij}^{\rm phot}\noindent\\[0.2cm]
&=& \delta C_{ij}^{\rm train} - \delta C_{ij}^{\rm phot}
\end{eqnarray}
where the second line trivially follows given that the true, underlying power
spectra are the same for the training and photometric galaxies. All of the
shear power spectra biases $\delta C$ can straightforwardly be evaluated from
Eq.~(\ref{eq:delta_Cmp}) by using the contamination coefficients for the
training and photometric samples, respectively.  Therefore, the effective error
in the power spectra is equal to the difference in the biases of the training
set (our {\it estimates} of the biases in the observable quantities) and the
photometric set (the actual biases in the observables).

\section{Probwts}\label{app:probwts}

In this subsection, we briefly review the weighting method\footnote{The
weights code is available at \url{ http://www.stanford.edu/~ccunha/nearest/}. 
The codes can also be obtained in the git repository {probwts} at \url{http://github.com}} 
of \cite{lim08} and \cite{cun09}.
We define the weight, $w$, of a galaxy in the spectroscopic training set as the
normalized ratio of the density of galaxies in the photometric sample to the
density of training-set galaxies around the given galaxy.  These densities are
calculated in a local neighborhood in the space of photometric observables,
e.g.\ multi-band magnitudes.  In this case, the DES {\it griz} magnitudes are
our observables.  The hypervolume used to estimate the density is set here to
be the Euclidean distance of the galaxy to its $N^{\rm th}$ nearest-neighbor in
the training set.  We set $N=2$, to derive the most localized estimates possible.

The weights can be used to estimate the redshift distribution of the
photometric sample using
\begin{equation}  
\nwei   = \sum_{\beta=1}^{N_{\rm T,tot}} w_\beta N(z_1<z_\beta<z_2)_{\rm T},
\label{eqn:Nzest}
\end{equation}
where the weighted sum is over all galaxies in the training set. \citet{lim08}
and \citet{cun09} show that this provides a nearly unbiased estimate of
the redshift distribution of the photometric sample, $N(z)_{\rm P}$, provided
the differences in the selection of the training and photometric samples are
solely done in the observable quantities used to calculate the weights.

\smallskip

\bibliographystyle{mn2e_new}
\bibliography{paper}

\end{document}